\documentclass[
aps,
prc,
twocolumn,
superscriptaddress,
nofootinbib,
amssymb,
amsmath,
amsfonts
]{revtex4-2}

\usepackage[dvipsnames]{xcolor}
\usepackage[utf8]{inputenc}
\usepackage{physics}  
\usepackage{graphicx}  
\graphicspath{{Graphics/}}  
\usepackage{booktabs}  
\usepackage{multirow}  
\usepackage{array}
\usepackage{tikz}
\usetikzlibrary{arrows.meta,automata,positioning,shadows,calc}
\usepackage{overpic}
\usepackage{hyperref}
\usepackage[capitalise]{cleveref}  

\newlength{\hoeheX}
\newlength{\breiteX}
\begin{document}
\title{Truncated Partial-Wave Analysis for $\eta$-photoproduction observables \\ via Bayesian Statistics}

\author{Philipp Kroenert}
\author{Yannick Wunderlich}
\author{Farah Afzal}
\author{Annika Thiel}
\affiliation{Helmholtz Institut für Strahlen- und Kernphysik, Universität Bonn, Germany}

\begin{abstract}
    A truncated partial-wave analysis is performed for $\eta$-photoproduction off the proton using the polarization observables $\sigma_0, \Sigma, T, E, F$ and $G$.
    Through this approach, model-independent estimates of the electromagnetic multipole parameters are calculated.
    Based on these estimates, predictions are made for polarization observables that have not yet been measured.
    These predictions identify promising future measurements that could resolve the inherent mathematical ambiguities within the results.
    Bayesian inference is combined for the first time with truncated partial-wave analysis, analyzing different truncation orders for six energy-bins near the $\eta p$-production threshold, i.e. $E^{lab}_{\gamma} \in [750, 1250]$~MeV.
\end{abstract}

\maketitle

\section{Introduction}
Baryon spectroscopy is an experimental technique to acquire a better understanding of the strong interaction and its fundamental theoretical description given by quantum chromodynamics.
Particles, for example pions, real photons as well as electrons \cite{THIEL2022103949}, are brought to collision with a nucleon.
With a sufficient high centre-of-mass energy, the nucleon can be excited to a resonant state, which is classified as a distinct particle with certain intrinsic properties.
Two well-established examples for baryon resonances are the Delta resonance $\Delta(1232)3/2^+$ and the Roper resonance $N(1440)1/2^+$ \cite{pdg}.
As such resonances are often formed and decay via the strong interaction, their proper lifetimes are rather short, for the above examples in the order of $10^{-24}$ s.
A direct detection of resonances with state-of-the-art detectors is not possible.
Instead, the analysis of the final state particles angular distributions using partial-wave analysis (PWA), allows to draw conclusions about the formation of the resonance and its inherent properties such as total angular momentum, mass, decay width and parity.
Up to the present day, single pseudoscalar meson photoproduction reactions are the experimentally most studied reactions in terms of baryon spectroscopy.
A comprehensive overview can be found in the recently published review on light baryon spectroscopy by Thiel et al. \cite{THIEL2022103949}.
The experimental data which are used as input to partial-wave analyses are called polarization observables.
In single pseudoscalar meson photoproduction, there are sixteen linearly independent measurable quantities.
Multiple facilities worldwide \cite{haensel1988european,WALCHER1990189,hillert2006bonn,rode2012continuous,YAMANAKA1992239} have contributed to a large database.
In addition, multiple PWA approaches \cite{PhysRevC.74.045205,anisovich2012properties,ronchen2018impact,DRECHSEL1999145,PhysRevC.54.2660,PhysRevC.88.035206} do exist for describing the data and extracting information about the resonant states.
The results of this paper are compared to the $K$-Matrix model of Bonn-Gatchina \cite{anisovich2012properties}, the dynamical coupled-channel approach of Jülich-Bonn \cite{ronchen2018impact} and the unitarized isobar-model of Eta-MAID \cite{DRECHSEL1999145}.
However, these approaches depend on an energy-dependent parameterization for the complex amplitudes \cite{THIEL2022103949}, leading to model-dependent outcomes.
For a detailed comparison of these three PWA approaches, the reader is advised to Refs. \cite{THIEL2022103949,anisovich2016impact}.
Resonant states can also be predicted in a purely mathematical manner via theory models based on quantum chromodynamics, such as quark models or Lattice quantum chromodynamics, see for example \cite{ronniger2011effects}.
However, theory models predicted significantly more states than are experimentally confirmed, predominantly in the higher-mass region, which is known as the missing-resonance problem \cite{THIEL2022103949}.
This unsolved issue motivates for further studies and the exploration of new approaches within this field of physics.
In this paper, a completely model-independent analysis approach, namely truncated partial-wave analysis (TPWA) \cite{Grushin,tiator2012towards,PhysRevC.89.055203,phdThesisYannick,wunderlich2017determining}, is employed.
This avoids the bias present in other PWA approaches.
In general, PWA as well as TPWA may exhibit mathematical ambiguities in their results, indicating that various solutions can effectively describe the same data points.
These ambiguities arise from the intrinsic mathematical nature of the problem.
As such, it is an essential step in any analysis using experimental data to check for potential ambiguities and evaluate their significance in comparison to each other.
Mathematical ambiguities in TPWA were first investigated by Omelaenko \cite{omelaenko1981ambiguities}.
A detailed treatment of the subject can be found in Refs. \cite{phdThesisYannick,PhysRevC.89.055203,wunderlich2017determining}.
The application of TPWA on experimental data ($\pi^0$-photoproduction off the proton for the first and second resonance region) was conducted in detail by Wunderlich \cite{phdThesisYannick} using the maximum likelihood method.
Among other things, the effect of measurement uncertainties on ambiguities was investigated.
This paper is the first to perform a TPWA using Bayesian inference.
Therefore, the results in this paper are given as distributions, as opposed to point estimates in previous PWA and TPWA approaches, allowing the uncertainty of an estimated parameter to be quantified with an unprecedented level of detail, which is of particular importance.
Through this approach it becomes possible to study the phase space in more detail and, by association, the structure of the above-mentioned ambiguities. It is even possible to discover a certain connectivity between different solutions, indicating problematic ambiguities.
The results of this paper comprises the estimation of complex electromagnetic multipole parameters for various maximal angular momenta $\ell_\text{max}$.
Based on these estimations, for the first time, model-independent predictions for unmeasured polarization observables are computed.
\\\\
The paper is structured as follows:
a concise introduction to Bayesian statistics is given in \cref{sec:introduction_bayesian_statistics}.
An outline of TPWA, hence the foundation of the employed model, is provided in \cref{sec:used_model}, followed by a discussion of the mathematical ambiguities.
The employed data sets are introduced and discussed in \cref{sec:database}, accompanied by the discussion of their systematic uncertainties and correlations between the used data points.
Within \cref{sec:posterior_distribution} the posterior distribution, centerpiece of the analyses, is introduced.
Finally, the results of TPWA examined via Bayesian inference are presented in \cref{sec:results}.

\section{Basics of Bayesian statistics}\label{sec:introduction_bayesian_statistics}
The fundamental equation of Bayesian statistics is Bayes' theorem \cite{bayes1763lii,gelman2013bayesian}:
\begin{equation}
    p(\vb*{\Theta} \mid \vb*{y}) = \frac{p(\vb*{y} \mid \vb*{\Theta}) \; p(\vb*{\Theta})}{\int p(\vb*{y} \mid \vb*{\Theta}) \; p(\vb*{\Theta}) \dd{\vb*{\Theta}}}. \label{eq::bayestheorem}
\end{equation}
Hereby, $\vb*{\Theta}$ denotes the parameters of the used model whereas $\vb*{y}$ stands for the employed data.
\\\\
The posterior distribution $p(\vb*{\Theta} \mid \vb*{y})$ is in general a multidimensional probability distribution reflecting the probability of the model given the data.
It consists of the likelihood distribution $p(\vb*{y} \mid \vb*{\Theta})$, comprising the data points and model predictions, and the prior distribution $p(\vb*{\Theta})$, which inhibits the current knowledge about the parameters of the model, before the data is taken into consideration.
The denominator in \cref{eq::bayestheorem} plays the role of a normalization factor and can be neglected within the computations of parameter estimation as it is constant for fixed $\vb*{y}$.
The definitions for the likelihood distribution and prior distributions employed in this paper can be found in \cref{sec:likelihoodDistribution,sec:prior_distribution}.
\\\\
The overall goal of each analysis is to scan the relevant regions of the posterior accurately.
From this, the parameter distributions can then be extracted, i.e. their marginal distributions\footnote{The marginal distribution of $\Theta_1$ with respect to the posterior distribution $p(\Theta_1,\Theta_2 \mid \vb*{y})$ is defined as $p(\Theta_1 \mid \vb*{y}) = \int \dd{\Theta_2} p(\Theta_1,\Theta_2 \mid \vb*{y})$. \cite{gelman2013bayesian}}.
    In general, the posterior is non-trivial and the integrals encountered in the derivation of the marginal distributions cannot be solved analytically.
    Instead, one can employ numerical methods, such as Markov chain Monte Carlo (MCMC) algorithms, in order to estimate the involved integrals.
    For instance, the Metropolis-Hastings \cite{metropolis1953equation, hastings1970monte} or the Hamiltonian Monte Carlo \cite{duane1987hybrid, neal2011mcmc} algorithm can be used, of which the latter one is applied in this work.
    The convergence of the Markov chains\footnote{``\textit{A sequence $X_1, X_2, \dots$ of random elements of some set is a Markov chain if the conditional distribution of $X_{n+1}$ given $X_1, \dots , X_n$ depends on $X_n$ only}.'' \cite[p. 2]{geyer2011introduction}.} can be monitored by convergence diagnostics such as the potential-scale-reduction statistic $\hat{R}$ \cite{gelman1992inference}, Monte Carlo standard error \cite{geyer2011introduction} (which depends on the effective sample size \cite{gelman2013bayesian}) and trace plots \cite{gelman2011inference}.
    \\\\
    To check the plausibility of the model under consideration, a posterior predictive check can be performed \cite{gelman2013bayesian}.
    Hereby, replicated data distributions $\vb*{y}^\text{rep}$ are generated using the sampled parameter distributions as input for the posterior distribution, while at the same time treating the data points as unknown parameters.
    In contrast to maximum likelihood or maximum a posteriori estimation, the marginal parameter estimates of Bayesian inference are given as distributions. This allows to quantify the uncertainty of a parameter with an unmatched level of detail.
    In addition, point estimates and the marginal parameter estimates of Bayesian inference differ in their underlying interpretation, making the latter an intriguing additional analysis approach.

    \section{Truncated partial-wave analysis}\label{sec:used_model}
    Within this section, the basic equations of TPWA for single pseudoscalar-meson photoproduction are outlined.
    For an in depth explanation, the reader is advised to Refs. \cite{phdThesisYannick, wunderlich2017determining}.
    \\\\
    Polarization observables are the measurable quantities of interest in single pseudoscalar-meson photoproduction.
    They are used as experimental input for a TPWA.
    In total there are sixteen polarization observables, which can be calculated by measuring differential cross sections under different polarization states.
    Three groups can be distinguished: the unpolarized differential cross section, three single-polarization observables and twelve double-polarization observables \cite{sandorfi2011determining}.
    A comprehensive list of the required polarization states for each observable is given in \cref{tab::polarization_configurations_observables} while a mathematical definition is given in \cref{sec:DiscreteAmbSixObservables}, \cref{tab:observable_definitions}.
    \begin{table}[h]
        \caption{This table collects the polarization configurations (beam, target, recoil) which allow to measure the sixteen polarization observables of pseudoscalar meson photoproduction. In the center-of-mass coordinate system, the unprimed coordinates are chosen as follows: $\hat{z}$-axis along incident photon beam direction and $\hat{y}$ perpendicular to the reaction plane $\hat{x}-\hat{z}$. The primed coordinates is a rotation of the unprimed coordinates such that the final state meson momentum points along the $\hat{z}'$-axis. The table is redrawn from Ref.~\cite{sandorfi2011determining}. A mathematical definition of the observables can be found in \cref{sec:DiscreteAmbSixObservables}, \cref{tab:observable_definitions}. \label{tab::polarization_configurations_observables}
        }
        \begin{ruledtabular}
            \begin{tabular}{ccc}
                \multirow{2}{*}{Observable} & Beam & Direction of target-/recoil- \\
                & polarization & nucleon polarization \\
                \midrule
                $\sigma_0$ & unpolarized & --- \\
                \midrule
                $\Sigma$ & linear & ---\\
                T & unpolarized & y\\
                P & unpolarized & y'\\
                \midrule
                H & linear & x \\
                P & linear & y \\
                G & linear & z \\
                F & circular & x\\
                E & circular & z\\
                \midrule
                $O_{x'}$ & linear & x'\\
                T & linear & y'\\
                $O_{z'}$ & linear & z'\\
                $C_{x'}$ & circular & x'\\
                $C_{z'}$ & circular & z'\\
                \midrule
                $T_{x'}$ & unpolarized & x, x' \\
                $L_{x'}$ & unpolarized & z, x' \\
                $\Sigma$ & unpolarized & y, y' \\
                $T_{z'}$ & unpolarized & x, z' \\
                $L_{z'}$ & unpolarized & z, z'
            \end{tabular}
        \end{ruledtabular}
    \end{table}
    \\\\
    The theoretical prediction of a profile function\footnote{The profile function $\check{\Omega}^{\alpha}\qty(W,\theta)$ of an observable $\Omega^{\alpha}\qty(W,\theta)$ is defined as $\check{\Omega}^{\alpha}\qty(W,\theta) := \sigma_0\qty(W,\theta) \cdot \Omega^{\alpha}\qty(W,\theta)$, where $\sigma_0$ is the unpolarized differential cross-section.} of a polarization observable depends on the energy $W$ as well as the scattering angle $\theta$ in the center-of-mass frame.
    It can be expressed as an expansion into the basis of associated Legendre polynomials $P^{\beta_{\alpha}}_{k}$ \cite{wunderlich2017determining}:
    \begin{equation}
        \check{\Omega}^{\alpha}_\text{theo}\qty(W,\theta) = \rho \sum^{2 \ell_\text{max} + \beta_{\alpha} + \gamma_{\alpha}}_{k=\beta_{\alpha}} \mathcal{A}^{\alpha}_{k}(W) \; P^{\beta_{\alpha}}_{k}\qty(\cos \theta). \label{eq:observable}
    \end{equation}
    \Cref{eq:observable} includes a kinematic phase-space factor $\rho$, angular expansion parameters $\beta_{\alpha}, \gamma_{\alpha}$, which are fixed parameters for each of the sixteen polarization observables of pseudoscalar-meson photoproduction, and energy dependent series coefficients $\mathcal{A}^{\alpha}_{k}$:
    \begin{equation}
        \mathcal{A}^{\alpha}_k(W) = \mathcal{M}^\dagger(W) \cdot \vb{\mathcal{C}}^\alpha_k \cdot \mathcal{M}(W). \label{eq:bilinear_product}
    \end{equation}
    Here, $\mathcal{M}$ denotes the complex multipole vector, which contains all participating multipoles involved for the truncation order $\ell_\text{max}$.
    A valid choice for the definition of this vector, by means of electromagnetic multipoles \cite{PhysRev.106.1345}, is:
    \begin{align}
        & \mathcal{M}(W) = [ \notag\\
        & \; \; E_{0+}(W), E_{1+}(W), M_{1+}(W), M_{1-}(W), \notag\\
        & \; \; E_{2+}(W), E_{2-}(W), M_{2+}(W), M_{2-}(W), \dots, \notag\\
        & \; \; E_{\ell_\text{max}+}(W), E_{\ell_\text{max}-}(W), M_{\ell_\text{max}+}(W), M_{\ell_\text{max}-}(W) \notag\\
        & ]. \label{eq:mulitpole_vector}
    \end{align}
    In addition, \cref{eq:bilinear_product} contains a complex $4\ell_\text{max} \cross 4\ell_\text{max}$ matrix $\mathcal{C}$ for each observable $\alpha$ and each summand $k$.
    Its general definition can be found in \cite{phdThesisYannick}\footnote{An overall factor of 1/2 is missing in the formula for $\mathcal{C}^\alpha_k$ in \cite{phdThesisYannick}.}.
    From these matrices one can not only read off the contributing partial-waves but also their interferences with each other \cite{wunderlich2017determining}.

    \Cref{eq:observable,eq:bilinear_product,eq:mulitpole_vector} imply:
    \begin{enumerate}
        \item The statistical analysis is performed for a single energy at a time.
        \item The polarization observable $\Omega^{\alpha}\qty(W, \theta)$ and the unpolarized differential cross-section $\sigma_0\qty(W, \theta)$ have to share the same energy- and angular-binning.
        \item The observables $\Omega^{\alpha}\qty(W,\theta)$ used within the TPWA have to share the same energy binning.
        \item As $\check{\Omega}^{\alpha}\qty(W,\theta)$ is an observable, i.e. a real number, the matrices $\vb{\mathcal{C}}^\alpha_k$ are hermitian.
        \item The bilinear form of $\mathcal{A}^{\alpha}_{k}$ gives rise to mathematical ambiguities, as certain transformations leave this quantity invariant.
    \end{enumerate}
    The last point is discussed in more detail in the following.

    \subsection{Ambiguities}\label{sec:multimodality}
    Ambiguities in PWA or TPWA refer to situations in which multiple configurations of the model parameters can describe the data points with similar levels of accuracy.
    This phenomenon is apparent in the reproduced data distributions in \cref{fig::reproduced_data_lmax_1,fig::reproduced_data_lmax_2} where the different colored distributions, corresponding to multiple ambiguities, nearly overlap.
    In the following discussion, various types of mathematical ambiguities are examined and it is concluded that only so-called accidental discrete ambiguities can appear in the results of this paper.
    \\\\
    The origin of the immanent mathematical ambiguities lies in the definition of the polarization observables.
    For photoproduction, they can be written in general as a bilinear product of the form \cite{PhysRevC.55.2054,PhysRevC.103.014607,PhysRevC.53.593}:
    \begin{equation}
        \check{\Omega}^\alpha(W, \theta) = \kappa \cdot b^\dagger(W, \theta) \; \Gamma^\alpha \; b(W, \theta),
    \end{equation}
    with a numerical prefactor $\kappa$, a vector $b$ of length $N_\text{A}$, containing the complex spin-amplitudes $b_i$, and a matrix $\Gamma^\alpha$ with dimensions $N_\text{A} \times N_\text{A}$.
    Certain transformations $T$ of the complex spin-amplitudes $b_i(W,\theta) \overset{T}{\longrightarrow} \tilde{b}_i(W,\theta)$ leave the bilinear product and thus the observable invariant.
    Hence, when all observables in a subset $\qty{\check{\Omega}^{\alpha_1}, \dots, \check{\Omega}^{\alpha_n}}$ are invariant under the same transformation, an ambiguity emerges \cite{PhysRevC.55.2054,phdThesisYannick}, as the experimental distinction between $b_i$ and $\tilde{b}_i$ is not possible any more.
    Such an ambiguity can be resolved by including a further observable $\check{\Omega}^{\alpha_k}$ into the subset, which is not invariant under the specific transformation \cite{PhysRevC.55.2054,phdThesisYannick}.
    There exists one special case of an ambiguity which cannot be resolved by including any further observables, namely the simultaneous rotation of all transversity amplitudes by the same (possibly energy and angle-dependent) phase: $b_i (W,\theta) \overset{T}{\longrightarrow} e^{i \phi (W,\theta)} b_{i} (W,\theta)$ (see \cite{PhysRevC.55.2054}).
    However, this continuous ambiguity can be ignored for the special case of a TPWA, since the angle-dependent part of the ambiguity is generally removed by the assumed truncation (see comments made in reference \cite{Wunderlich:2017dby}), and the energy-dependent part is fixed by imposing certain phase-conventions for the multipoles.
    The formalism for the remaining relevant discrete ambiguities in a TPWA is outline briefly in the following.
    For more information about discrete as well as continuous ambiguities in the case of the complete experiment analysis, see the paper of Chiang and Tabakin \cite{PhysRevC.55.2054}.
    \\\\
    As shown by Omelaenko \cite{omelaenko1981ambiguities,phdThesisYannick}, in a TPWA (truncated at some finite~$\ell_{\text{max}} \geq 1$) the complex spin-amplitudes can be expressed (up to kinematical prefactors) as a finite product of irreducible polynomials:
    \begin{align}
        b_1 (W, \theta) &\propto \prod_{k = 1}^{2\ell_\text{max}} \qty(\tan \frac{\theta}{2} + \beta_k(W)), \label{eq:b1LinFactDecomp} \\
        b_2 (W, \theta) &\propto \prod_{k = 1}^{2\ell_\text{max}} \qty(\tan \frac{\theta}{2} - \beta_k(W)), \label{eq:b2LinFactDecomp} \\
        b_3 (W, \theta) &\propto \prod_{k = 1}^{2\ell_\text{max}} \qty(\tan \frac{\theta}{2} + \alpha_k(W)), \label{eq:b3LinFactDecomp} \\
        b_4 (W, \theta) &\propto \prod_{k = 1}^{2\ell_\text{max}} \qty(\tan \frac{\theta}{2} - \alpha_k(W)), \label{eq:b4LinFactDecomp}
    \end{align}
    with the complex roots $\alpha_k(W)$ and $\beta_k(W)$, which are in essence equivalent to multipoles.
    It can be shown \cite{omelaenko1981ambiguities,PhysRevC.89.055203,PhysRevC.95.015206} that the special case where $\tan\frac{\theta}{2} = 0$ implies a direct connection between the roots:
    \begin{equation}
        \prod^{2\ell_\text{max}}_{i=1} \alpha_i(W) = \prod^{2\ell_\text{max}}_{j=1} \beta_j(W). \label{eq:consistency}
    \end{equation}
    All transformations $T$ which correspond to a discrete ambiguity of the four group~$\mathcal{S}$ observables~$\left\{ \sigma_{0}, \check{\Sigma}, \check{T}, \check{P} \right\}$, must also satisfy \cref{eq:consistency}, which allows to rule out a major part of the maximal possible $4^{2\ell_\text{max}}$\cite{PhysRevC.89.055203} discrete ambiguity-transformations from the beginning.
    The so-called~’double ambiguity’~\cite{omelaenko1981ambiguities,PhysRevC.89.055203}, which corresponds to the simultaneous complex conjugation of all root, automatically preserves the constraint in \cref{eq:consistency}.
    \\\\
    Unfortunately, there can also occur so-called accidental ambiguities.
    These emerge when any discrete ambiguity other than the double ambiguity of all roots approximately fulfills \cref{eq:consistency} \cite{PhysRevC.89.055203}.
    The accidental ambiguities as well as the double ambiguity can in principle be resolved by including further observables into the analysis apart from the four group~$\mathcal{S}$ observables.
    Candidates for observables capable of resolving the above-mentioned discrete ambiguities would be either~$\check{F}$, $\check{G}$ or any of the $\mathcal{BR}$- and $\mathcal{TR}$-type observables.
    \\\\
    The accidental ambiguities cannot be avoided for analyses of real data due to their abundance~(i.e.~$4^{2 \ell_{\text{max}}} - 2$ possible candidates exist for such ambiguities), and they will show up as modes within the posterior distribution and thus in the marginal parameter distributions.
    \\\\
    In contrast to the discrete ambiguities described above, there can also exist so-called continuous ambiguities in the TPWA (in addition to the above-mentioned simultaneous phase-rotation of all transversity amplitudes, which has been ruled out), which exist on continuously connected regions within the multipole parameter-space~\cite{phdThesisYannick}.
    These ambiguities can occur in case an insufficiently small set of observables is analyzed, and they manifest as plateau-like structures (with possibly rounded edges) in the marginalized posterior-distributions, as opposed to the peak-like structures (or modes) originating from discrete ambiguities.
    The set of six observables analyzed in this work (see \cref{sec:database}) is large enough to avoid such continuous ambiguities.
    \\\\
    For more information about discrete ambiguities in TPWA, the paper by Omelaenko \cite{omelaenko1981ambiguities} and especially the subsequent work \cite{PhysRevC.89.055203} is recommended.
    The proof of the completeness of the set of six observables analyzed in this work (\cref{sec:database}) in the idealized case of an 'exact' TPWA~\footnote{Accidental ambiguities can be disregarded for this rather academic scenario~\cite{phdThesisYannick}.} proceeds a little bit different compared to the work by Omelaenko~\cite{omelaenko1981ambiguities}.
    The proof is outlined in some detail in \cref{sec:DiscreteAmbSixObservables}.
    \\\\
    Summarizing, accidental discrete ambiguities will likely be present within TPWA performed on real data, resulting in a multimodal likelihood and posterior distribution.

    \section{Discussion of the used database}\label{sec:database}
\begin{figure}
        \includegraphics[width=\linewidth]{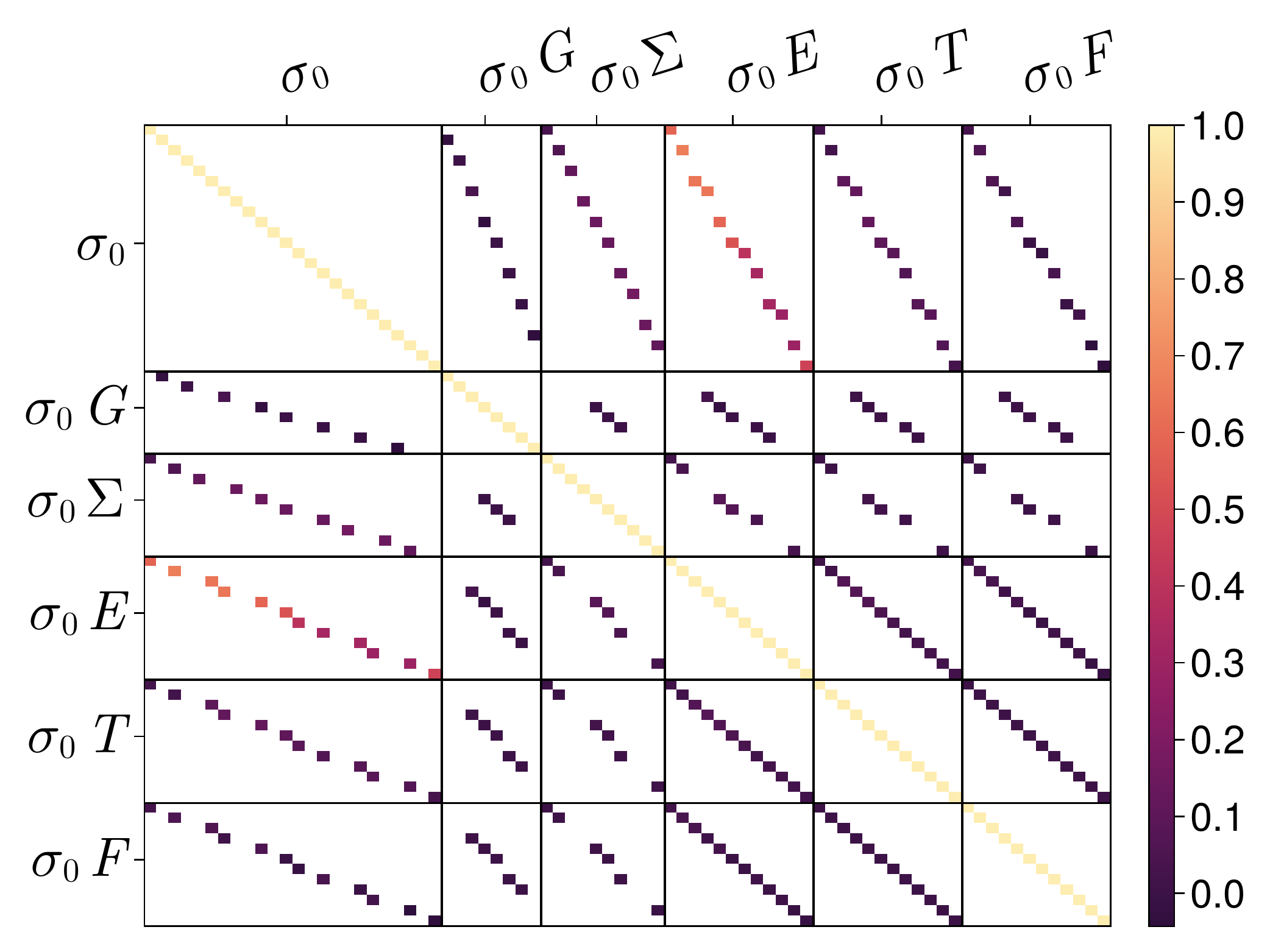}
        \caption{Example for a correlation matrix. The correlations between the data points of the unpolarized differential cross-section $\sigma_0$ and the used profile functions, as well as the correlation between the profile functions themselves, is shown for $E^\text{lab}_\gamma=750$~MeV.
        Each square represents a certain data point.
        The color encodes the correlation strength ranging from $-1$ (darker colors) to $+1$ (lighter colors).
        \label{fig:correlationMatrix750MeV}
        }
    \end{figure}

\begin{figure}
        \includegraphics[width=\linewidth]{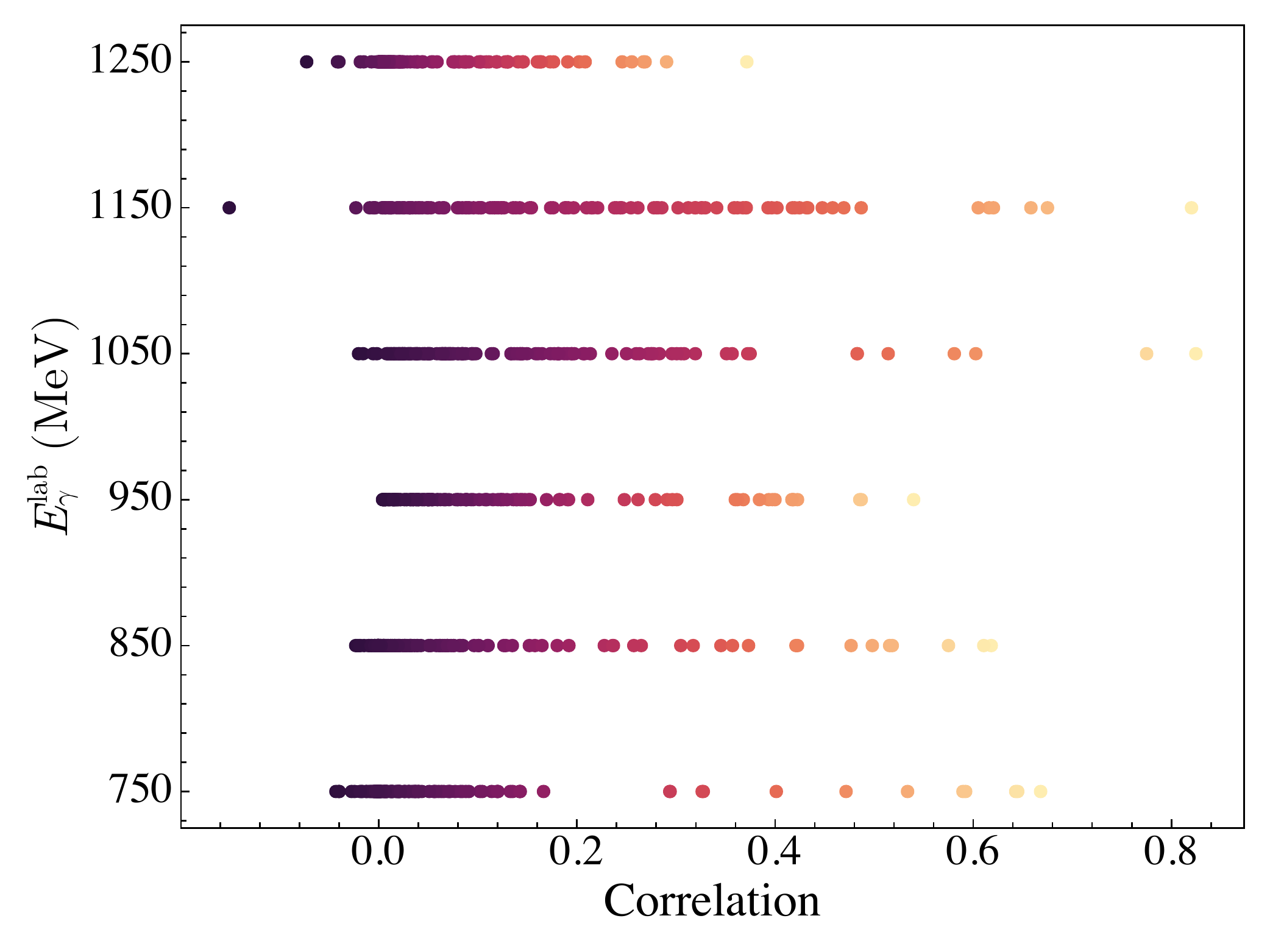}
        \caption{Unique correlation matrix element values as a function of the lab frame energy.
        The color encodes the correlation strength ranging from $-1$ (darker colors) to $+1$ (lighter colors).
        \label{fig:correlations_all_energies}
        }
    \end{figure}

    A review of the currently available database on polarization observables for the reaction $\gamma p \rightarrow \eta p$ can be found in \cite{THIEL2022103949}.
    In order to cover the largest possible energy range and to resolve discrete mathematical ambiguities, the TPWA is performed using the six polarization observables $\sigma_0$ \cite{PhysRevLett.118.212001}, $\Sigma$ \cite{bartalini2007measurement}, $T$  \cite{akondi2014measurement}, $E$ \cite{afzal2019measurement}, $F$ \cite{akondi2014measurement} and $G$ \cite{muller2020new}.
    This choice of observables indeed resolve the discrete ambiguities of TPWA, as shown in \cref{sec:DiscreteAmbSixObservables}.
    \\\\
    An overview of the data is given in \cref{tab:used_data} and a visualization of the phase-space coverage of the individual data sets can be found in \cref{sec:data_coverage}, \cref{fig:energy_angular_coverage}.
    \begin{table*}[t]
        \caption{Information on the experimental data, given as dimensionless asymmetries, used for the TPWA of $\gamma p \rightarrow \eta p$. Energy and angular ranges are written as intervals. \label{tab:used_data}}
        \centering
        \begin{ruledtabular}
            \begin{tabular}{ccccll}
                Observable & Number of data points & $E^\text{lab}_{\gamma}$ / MeV & $\cos(\theta)$ & Facility & References\\
                \midrule
                $\sigma_0$ & 5736 & $[723, 1571]$ & $[-0.958, 0.958]$ & MAMI & Kashevarov et al. \cite{PhysRevLett.118.212001}\\
                \midrule
                $T,F$ & 144 & $[725, 1350]$ & $[-0.917, 0.917]$ & MAMI & Akondi et al. \cite{akondi2014measurement}\\
                \midrule
                $\Sigma$ & 140 & $[761, 1472]$ & $[-0.946, 0.815]$ & GRAAL & Bartalini et al. \cite{bartalini2007measurement}\\
                \midrule
                $E$ & 84 &$[750, 1350]$ & $[-0.917, 0.917]$ & MAMI & Afzal et al. \cite{afzal2019measurement,observableE2}\\
                \midrule
                $G$ & 47 & $[750, 1250]$ & $[-0.889, 0.667]$ & CBELSA/TAPS & Müller et al. \cite{muller2020new}\\
            \end{tabular}
        \end{ruledtabular}
    \end{table*}
    The available energies for the TPWA are determined by the observable with the lowest statistics \cite{sandorfi2011determining, phdThesisYannick}, which in this case is the observable $G$.
    In total six energy-bins are available, starting near the $\eta p$-photoproduction threshold at $E_\gamma^\text{lab} = 750$~MeV up to 1250 MeV, in 100 MeV steps.
    \\\\
    As TPWA is a single-energy regression, the energy binning of each observable has to be shifted to that of $G$.
    The procedure is described in \cite{phdThesisYannick}.
    The advantage of this method is that no new, i.e. experimentally unobserved, data points have to be constructed, for example via interpolation.
    \\\\
    However, none of the observables are given as profile-functions which are needed for the TPWA, see \cref{eq:observable}.
    Thus, the angular distribution of $\sigma_0$ has to be adjusted for each observable, in order to multiply both.
    This is not an issue, since the very precise MAMI $\sigma_0$-dataset \cite{PhysRevLett.118.212001} covers a large angular range $[-0.958, 0.958]$ with a small step size $\sim 0.083$ in all available energies.
    \\\\
    The data discussed in \cref{sec:database} do not only have statistical- but also systematic uncertainties.
    The latter ones originate primarily from the determination of the polarization degree of the photon beam and the target nucleon, the dilution factor\footnote{The dilution factor is the ratio of polarizable free protons to all nuclei in the used target material.} as well as the background subtraction procedure \cite{muller2020new,afzal2019measurement,akondi2014measurement,PhysRevLett.118.212001,bartalini2007measurement}.
    \\\\
    In principle, each data point has its own systematic uncertainty.
    However, there is no generally accepted method to model the systematic uncertainty for each data point separately.
    Instead, the contributions to the systematic uncertainty, which are constant over the whole angular range, are determined for each data set.
    Then, the same systematic uncertainty is used for each data point within a data set.
    \\\\
    The contributions split up into the ``\textit{general systematic uncertainty}'' ($\sigma_0$: $4\%$ \cite[p. 5]{PhysRevLett.118.212001}), the degree of photon beam polarization (F: $2\%$ \cite{akondi2014measurement}, E: $2.7\%$ \cite{afzal2019measurement}, G: $5\%$ \cite{muller2020new}) and the degree of target polarization (T,F: $4\%$ \cite{akondi2014measurement}, E: $2.8\%$ \cite{afzal2019measurement}, G: $2\%$ \cite{muller2020new}).
    The authors of the polarization observable $\Sigma$ added the statistical- and systematic uncertainty in quadrature for each data point \cite{bartalini2007measurement}.
    Thus, their systematic uncertainty can not be modeled separately within this paper.
    \\\\
    The individual systematic contributions within a data set are combined in a conservative way.
    A worst-case scenario approach is employed, based on the formulas used to calculate the polarization observables, as given in the papers.
    In comparison with the 'standard' procedure of adding the different contributions in quadrature, there are two main advantages:
    1) The functional dependence is taken into account without the need to make an assumption about the distribution of the individual contributions.
    2) The worst-case scenario covers the maximum/minimum impact of the systematic uncertainties, and everything in between.
    \\\\
    As an illustrative example, suppose an observable $A$ which depends reciprocal on the degree of polarization of the photon beam $p_\gamma$ and target $p_\text{t}$, each with their own relative systematic uncertainty $\Delta^{p_\gamma}_\text{sys}$ and $\Delta^{p_\text{t}}_\text{sys}$, respectively.
    Then the combined, relative systematic uncertainty of $A$ would be:
    \begin{align}
        \Delta^{A}_\text{sys} = \max \Bigl( & \qty|1 - (1 + \Delta^{p_\gamma}_\text{sys})^{-1} \cdot (1 + \Delta^{p_\text{t}}_\text{sys})^{-1}|, \notag\\
        & \qty|1 - (1 - \Delta^{p_\gamma}_\text{sys})^{-1} \cdot (1 - \Delta^{p_\text{t}}_\text{sys})^{-1}|\Bigr).
    \end{align}
    With input taken from the references, corresponding to the respective data sets \cite{PhysRevLett.118.212001,bartalini2007measurement,muller2020new,afzal2019measurement,akondi2014measurement}, the outlined approach results in:
    $\Delta^{\sigma_0}_\text{sys} = 4.0\%$,
    $\Delta^\text{G}_\text{sys} = 7.4\%$,
    $\Delta^\text{E}_\text{sys} = 5.7\%$,
    $\Delta^\text{T}_\text{sys} = 4.2\%$,
    $\Delta^\text{F}_\text{sys} = 6.3\%$.
    \\\\
    Due to the calculation of the profile functions, the systematic uncertainty of both data sets have to be combined as well:
    \begin{align}
        \Delta^{\check{A}}_\text{sys} = \max \Bigl( & \qty|1 - (1 + \Delta^{A}_\text{sys}) \cdot (1 + \Delta^{\sigma_0}_\text{sys})|, \notag\\
        & \qty|1 - (1 - \Delta^{A}_\text{sys}) \cdot (1 - \Delta^{\sigma_0}_\text{sys})| \Bigr).
    \end{align}
    Thus, the relative systematic uncertainties for the profile functions are:
    $\Delta^{\sigma_0}_\text{sys} = 4.0\%$,
    $\Delta^{\check{G}}_\text{sys} = 11.7\%$,
    $\Delta^{\check{E}}_\text{sys} = 10.0\%$,
    $\Delta^{\check{T}}_\text{sys} = 8.3\%$,
    $\Delta^{\check{F}}_\text{sys} = 10.5\%$.
    The incorporation of the systematic uncertainties into the statistical model is described in more detail in \cref{sec:posterior_distribution}.
    \\\\
    Furthermore, the calculation of the profile functions introduces a correlation between the unpolarized differential cross-section and the profile functions, as well as among the profile functions themselves.
    Since certain values of $\sigma_0(W, \theta)$ were used to calculate $\check{\Omega}^\alpha(W, \theta)$, correlations were introduced between certain data points of both observables.
    Moreover, the same value of $\sigma_0(W, \theta)$ might be used to calculate data points of different profile functions.
    \\\\
    The relevance of these correlations can be estimated via the Pearson correlation coefficient \cite{definition_correlation}, see \cref{eq:correlation1,eq:correlation2} in \cref{sec:definitionCorrelation}.
    The measured values of the polarization observables are used as expectation values and the corresponding squared statistical uncertainties as the variances.
    An example for a correlation matrix is shown in \cref{fig:correlationMatrix750MeV}.
    The correlations are quite small, with absolute values below $\sim 0.17$, but typically in the order of $10^{-2} - 10^{-3}$.
    An exception is the significantly higher correlation between $\sigma_0$ and $\sigma_0 \cdot E$, with minimal and maximal values of $\sim 0.29$ and $\sim 0.67$, respectively.
    This can be explained by the similar definition of the coefficients $\mathcal{A}^{\alpha}_k(W)$ of $\sigma_0$ and $\sigma_0\cdot E$.
    Both having sensitivity to almost the exact same interference terms of multipoles, albeit with different strengths (see Ref.~\cite{wunderlich2017determining}).
    The magnitude of the correlation matrix elements as a function of the energy can be seen in \cref{fig:correlations_all_energies}.
    The corresponding covariance matrix, which is used to construct the likelihood distribution in \cref{sec:likelihoodDistribution}, can be estimated via \cref{eq:covariance1,eq:covariance2} in \cref{sec:definitionCorrelation}.

    \section{The posterior distribution}\label{sec:posterior_distribution}
    It is assumed that the utilized profile functions, constructed from the polarization observables, follow a normal distribution.
    The validity of this assumption is extensively discussed in \cref{sec:assumptions}.
    However, the profile functions are correlated with the unpolarized differential cross-section, as well as among themselves, see \cref{sec:database}.
    This dependence is modeled within the likelihood distribution using a covariance matrix.
    In favor of a compact representation, the functional dependencies are not shown explicitly in the subsequent equations.

    \subsection{Likelihood distribution}\label{sec:likelihoodDistribution}
    Combining the results of \cref{sec:database,sec:assumptions}, the conditional likelihood distribution, for each of the analyzed energies, can be formulated as an $N$-dimensional multivariate Gaussian distribution:
    \begin{align}
        p(\vb*{y}, \vb*{x} \mid \vb*{\Theta}, \vb*{\kappa}) &= \mathcal{N}(\vb*{\mu}, \vb*{\Lambda}) \notag \\
        &= \frac{\exp(-\frac{1}{2} \qty(\vb*{y} - \vb*{\mu})^\mathrm{T} \vb*{\Lambda}^{-1}\qty(\vb*{y} - \vb*{\mu}))}{\sqrt{(2\pi)^N |\vb*{\Lambda}|}}.\label{eq:joint_likelihood}
    \end{align}
    Hereby, the vectors $\vb*{y},\vb*{x} \in \mathbb{R}^N$ contain the entirety of the $N \in \mathbb{N}$ utilized profile function data points and the corresponding $\cos(\theta)$-values at which they were measured, respectively:
    \begin{align}
        \vb*{y} &= [\vb*{y}^{\sigma_0}, \vb*{y}^{\check{G}}, \vb*{y}^{\check{\Sigma}}, \vb*{y}^{\check{E}},\vb*{y}^{\check{T}}, \vb*{y}^{\check{F}}], \\
        \vb*{x} &= [\vb*{x}^{\sigma_0}, \vb*{x}^{\check{G}}, \vb*{x}^{\check{\Sigma}}, \vb*{x}^{\check{E}}, \vb*{x}^{\check{T}}, \vb*{x}^{\check{F}}].
    \end{align}
    The parameters of the model can be divided into two groups.
    On the one hand, the real- and imaginary parts of multipoles, i.e. \cref{eq:mulitpole_vector}, denoted by $\vb*{\Theta} \in \mathbb{R}^{8\ell_\text{max}-1}$ are used to model the underlying physical process.
    On the other hand, the parameters $\vb*{\kappa} \in \mathbb{R}^5$ which are used to model the systematic uncertainties of the involved data sets:
    \begin{equation}
        \vb*{\kappa} = [\kappa^{\sigma_0}, \kappa^{\check{G}}, \kappa^{\check{E}}, \kappa^{\check{T}}, \kappa^{\check{F}}].
    \end{equation}
    The multivariate normal distribution in \cref{eq:joint_likelihood} is constructed with the model predictions $\vb*{\mu} \in \mathbb{R}^N$ for the expectations of $\vb*{y}$:
    \begin{align}
        \vb*{\mu}(\vb*{\Theta}, \vb*{\kappa}, \vb*{x}) = [
        &\kappa^{\sigma_0} \cdot \vb*{\mu}^{\sigma_0}, \kappa^{\check{G}} \cdot \vb*{\mu}^{\check{G}}, 1 \cdot \vb*{\mu}^{\check{\Sigma}}, \notag\\
        &\kappa^{\check{E}} \cdot \vb*{\mu}^{\check{E}}, \kappa^{\check{T}} \cdot \vb*{\mu}^{\check{T}}, \kappa^{\check{F}} \cdot \vb*{\mu}^{\check{F}}
        ].
    \end{align}
    The $\vb*{\mu}^\alpha(\vb*{\Theta}, \vb*{x}^\alpha)$ are the model predictions for the individual profile functions, i.e. \cref{eq:observable}.
    Hence, in order to model the systematic uncertainties, one additional parameter per relevant data set is introduced and multiplied with the corresponding theoretical prediction for the profile function.
    Thus, the model gets additional degrees of freedom to adjust for possible systematic uncertainties.
    However, these parameters are restricted to physical meaningful bounds, further discussed in \cref{sec:prior_distribution}.
    As explained in \cref{sec:database}, the systematic uncertainty of the polarization observable $\Sigma$ can not be modeled.
    \\\\
    Finally, there is the covariance matrix $\vb*{\Lambda} \in \mathbb{R}^{N \times N}$.
    Its off-diagonal terms are not identical, and therefore the data-pairs are not exchangeable\footnote{If the joint probability density function $p(\vb*{y}, \vb*{x} | \vb*{\Theta}, \vb*{\kappa})$ is invariant under permutations of the data-pairs $(y,x)_i$, then the data-pairs are said to be exchangeable \cite{gelman2013bayesian,o2009exchangeability}.}.

    \subsection{Prior distribution}\label{sec:prior_distribution}
    The priors for the multipole parameters are chosen as uniform priors with bounds corresponding to the physically allowed ranges of the parameters (see \cite{phdThesisYannick}).
    Thus, the priors incorporate physical knowledge while being uninformative compared to the likelihood distribution.
    \\\\
    In principle a uniform prior for the systematic parameters would be reasonable.
    However, in this case the hard boundaries in the parameter space lead to numerical issues.
    Thus, the prior distributions for the scaling parameters $\vb*{\kappa}$ are assumed to be normal distributed, and centered around the value one.
    The standard deviation is chosen such that\footnote{This can be calculated by solving numerically the following equation for the standard deviation $\sigma$:
    \begin{equation*}
        \int_{-\infty}^{1-\Delta^\alpha_\text{sys}} \frac{\exp(-\frac{1}{2}\qty(\frac{x-1}{\sigma})^2)}{\sigma \sqrt{2\pi}} \dd{x} = \frac{1-0.99}{2}.
    \end{equation*}
    }
    $99\%$ of the distribution are within the range $1 \pm \Delta^{\alpha}_\text{sys}$, which results in (rounded to five digits):
    \begin{align}
        \kappa^{\sigma_0} &\sim \mathcal{N}(1, 0.01552), \label{eq:prior1}\\
        \kappa^{\check{G}} &\sim \mathcal{N}(1, 0.04542), \label{eq:prior2}\\
        \kappa^{\check{E}} &\sim \mathcal{N}(1, 0.03882), \label{eq:prior3}\\
        \kappa^{\check{T}} &\sim \mathcal{N}(1, 0.03222), \label{eq:prior4}\\
        \kappa^{\check{F}} &\sim \mathcal{N}(1, 0.04076). \label{eq:prior5}
    \end{align}
    This choice is in accordance with the conservative combination of the systematic uncertainties as discussed in \cref{sec:database}.
    The treatment of systematic errors within this paper is similar to that in Refs. \cite{PhysRevC.93.065205,rnchen2022light, sandorfi2011determining}.

    \section{Results} \label{sec:results}
    Bayesian inference was utilized to extract the electromagnetic multipole parameters, as introduced in \cref{eq:mulitpole_vector}, for the reaction $\gamma p \rightarrow \eta p$ at energies $E^\text{lab}_\gamma = [750, 850, 950, 1050, 1150, 1250]$~MeV and truncation orders $\ell_\text{max}=1,2$ through truncated partial-wave analysis.
    The procedures involved are detailed in \cref{sec:analysis_steps}.
    \\\\
    Highly multimodal posterior distributions were encountered, necessitating an adaptation of the typical MCMC convergence diagnostic workflow.
    The adjusted procedure is detailed in \cref{sec:mcmcConvergence}.
    By studying the reproduced data distributions for the various truncation orders, an indication of $N^*$-resonances in the energy range from $950$ to $1050$~MeV is observed.
    Finally, utilizing the electromagnetic multipole parameter estimates, predictions were calculated\footnote{To get from the profile functions to the dimensionless polarization observables, the predicted distribution is divided by a certain $\sigma_0$-value, corresponding to the $\cos(\theta)$-value at which the prediction were calculated.} for the polarization observables $H$, $P$, as well as those of group $\mathcal{BR}$ and $\mathcal{TR}$.
    Hence, this includes eight polarization observables that are yet to be measured.
    The distributions of the multipole parameters, the reproduced and predicted data are presented alongside the values of EtaMAID2018 \cite{Tiator:2018heh}, BnGa-2019 \cite{muller2020new} and JüBo-2022 \cite{rnchen2022light}.
    \\\\
    The presentation of the multipole parameter results is quite detailed and deserves an explanation.
    The top part shows the solutions found via Monte Carlo maximum a posteriori estimation and their corresponding $\chi^2/ndf$ values, together with the $1\sigma$-uncertainty (see \cref{section:mcmle}).
    The middle part shows the marginal-parameter distributions obtained via Bayesian inference, as explained in \cref{sec:posterior_distribution,sec:posteriorSampling}.
    For a better comparison of the two approaches for $\ell_\text{max}=1$, the $[0.16, 0.5, 0.84]$-quantiles of the distributions, corresponding to the median of the distribution and the $1\sigma$-uncertainty boundaries, are drawn as dashed lines through all parts of the figure.
    Whereas, for $\ell_\text{max}=2$ a solid vertical line is drawn for each peak of the multimodal distribution, i.e. the most probable values.
    The bottom part of the figure is a contour plot of the log posterior density distribution and the corresponding marginal-parameter distribution.
    The outermost contour line is at $1\%$ of the maximum altitude, each subsequent line represents an $11\%$ increase.
    It is assumed that a log posterior distribution centered around a higher log posterior value, corresponds to more likely parameter values, as this solution contributes more probability mass to the posterior.
    Each solution group is drawn in a different color and each peak of a distribution corresponds to an accidental ambiguity.
    The color for a specific solution group is consistent between the shown figures (MCMC convergence- , multipole-, predictive performance plots, etc.) for a certain energy and truncation order.
    This means one can monitor the behavior of a specific ambiguity, ranging from the MCMC diagnostic plots in \cref{fig:diagnostics_lmax2} to the multipole plots in \cref{fig::multipole_plots_lmax_2_e0p,fig::multipole_plots_lmax_2_m2p_2}, to the reproduced data plots in \cref{fig::reproduced_data_lmax_2}, up to the predicted data distribution plots in \cref{fig::all_predicted_distributions_lmax_2}.
    The performed analyses showed, that Bayesian inference gives more insight into the relevance of ambiguities, due to the Hamiltonian Monte Carlo algorithm.
    When multiple chains sample consistently multiple marginal modes together, this is a sign of a problematic ambiguity, as they tend to have comparable log posterior densities.
    As an example, consider the multipole solution for $\mathrm{Re}(M_{2+})$ at $750 \; \mathrm{MeV}$ in \cref{fig::multipole_plots_lmax_2_e0p}.
    This is an advantage over the maximum likelihood approach.
    \\\\
    Within the following discussion of the results a representative selection of figures is shown.
    All parameter figures, for all analyzed energies and truncation orders can be found in the supplementary material Ref.~\cite{supplementMaterial}.

    \subsection{Impact of accidental ambiguities on the results}
\begin{figure*}[p]
        \includegraphics[width=\textwidth]{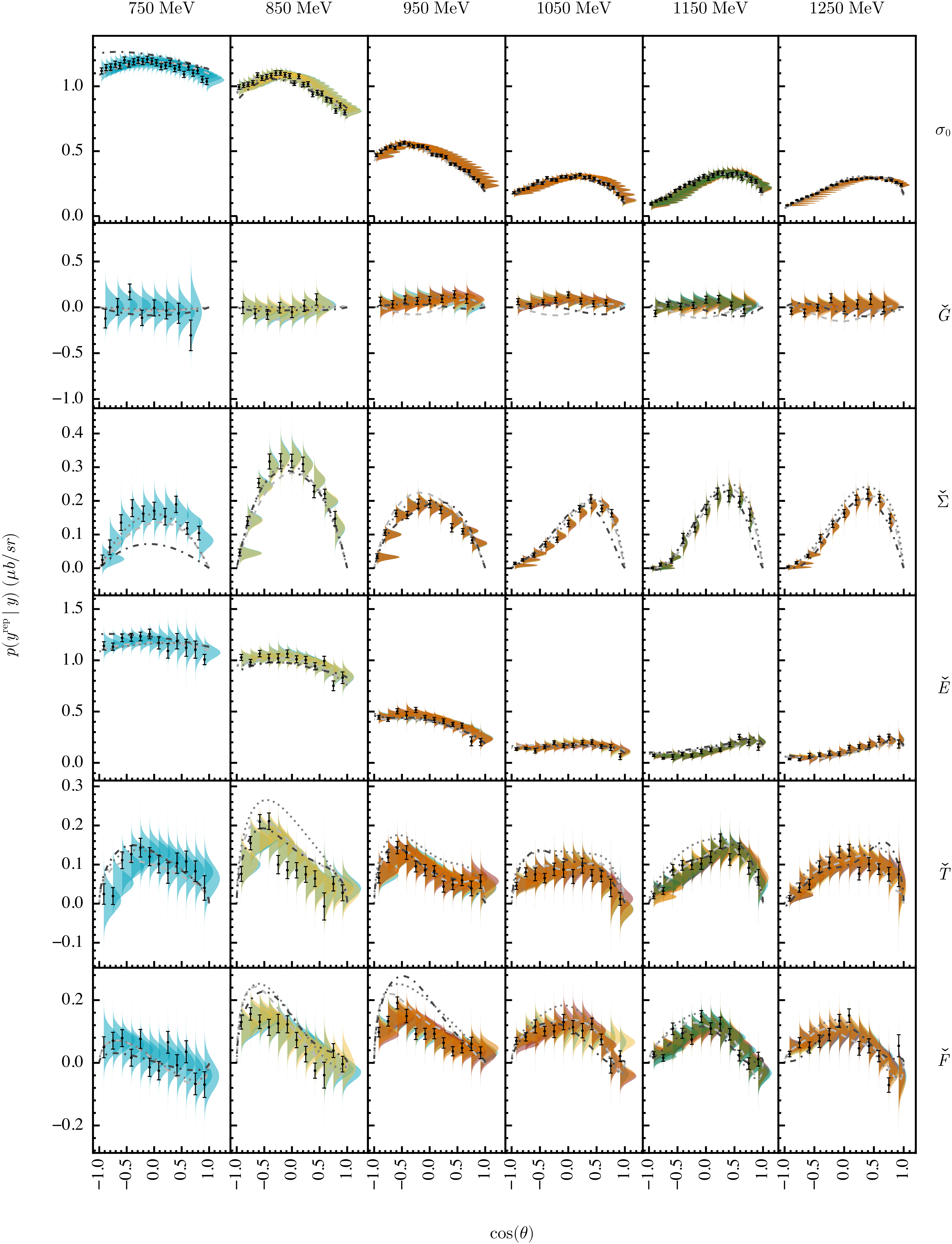}
        \caption{Posterior predictive check for the profile functions $\sigma_0, \check{G}, \check{\Sigma}, \check{E}, \check{T}$ and $\check{F}$, for truncation order $\ell_\text{max} = 2$ and energy-bins $E^\text{lab}_\gamma=\qty[750,850,950,1050,1150,1250]$~MeV. The reproduced data distributions for the different solutions are shown together with the original data with statistical uncertainties as black points. Each solution group is drawn in a different color and each peak of a distribution corresponds to an accidental ambiguity. In addition, the corresponding values from EtaMAID2018 \cite{Tiator:2018heh} (dashed line), BnGa-2019 \cite{muller2020new} (dotted line) and JüBo-2022 \cite{rnchen2022light} (dash-dotted line) are shown as well. \label{fig::reproduced_data_lmax_2}}
    \end{figure*}

    As discussed in \cref{sec:multimodality}, accidental ambiguities appear in the results of the marginal multipole parameter distributions, which subsequently manifest in both the replicated and predicted data distributions, as well as in the marginal systematic parameter distributions.
    The ambiguities are apparent as differently colored distributions, where each peak of a distribution corresponds to an accidental ambiguity.
    See for example \cref{fig::reproduced_data_lmax_2,fig::multipole_plots_lmax_2_e0p,fig::multipole_plots_lmax_2_m2p_2,fig::all_predicted_distributions_lmax_2}.
    As expected, all accidental ambiguities can replicate the original data points.
    The corresponding reproduced data distributions are nearly identical, as illustrated in \cref{fig::reproduced_data_lmax_2}.
    The impact of ambiguities on predicted data distributions and what can be learn from it is discussed in \cref{sec::predicted_data}.

    \subsection{Choice of the truncation order}
\begin{figure*}[p]
        \includegraphics[width=\textwidth, height=\textheight, keepaspectratio]{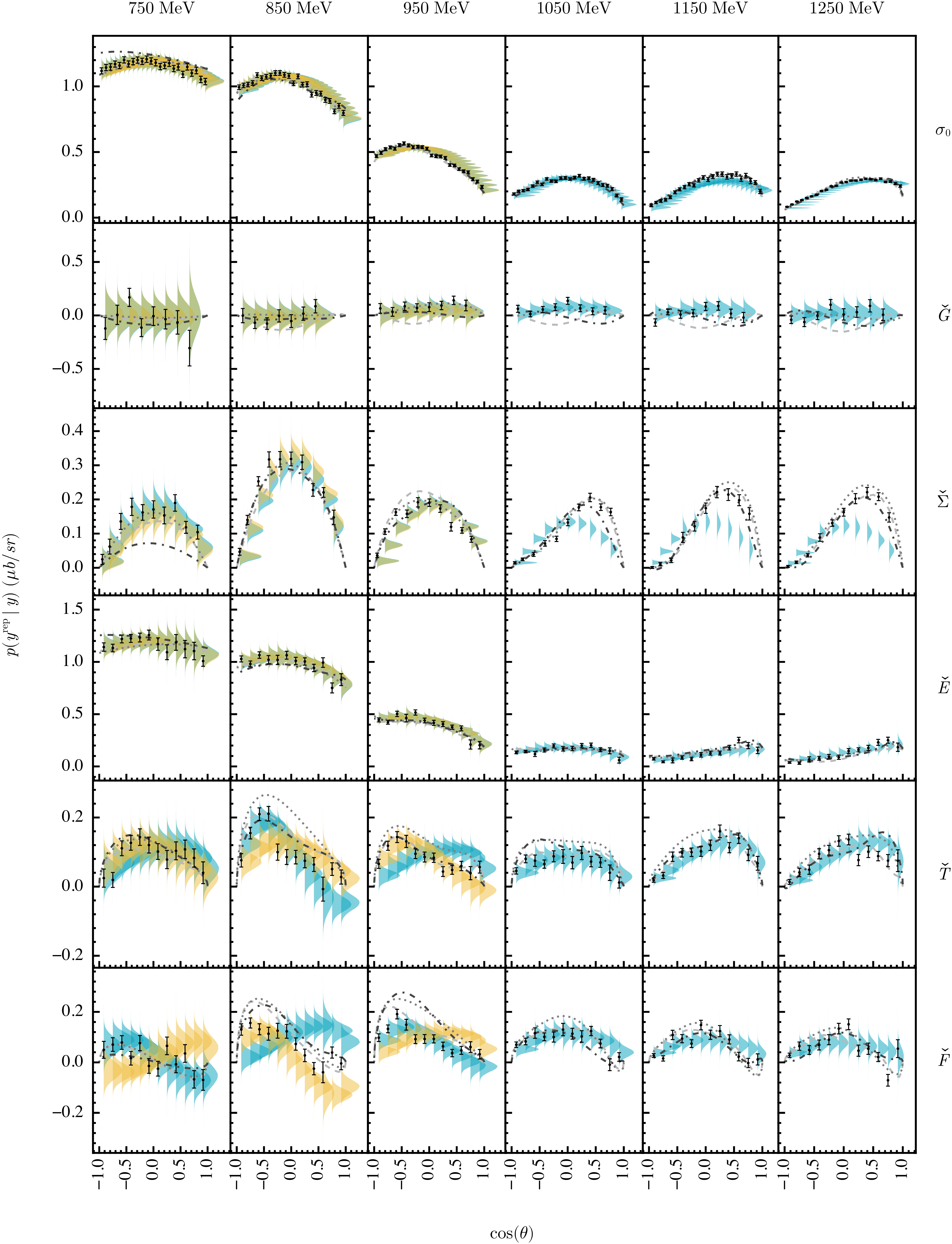}
        \caption{Posterior predictive check for the profile functions $\sigma_0, \check{G}, \check{\Sigma}, \check{E}, \check{T}$ and $\check{F}$, for truncation order $\ell_\text{max} = 1$ and energy-bins $E^\text{lab}_\gamma=\qty[750,850,950,1050,1150,1250]$~MeV. The reproduced data distributions for the different solutions are shown together with the original data with statistical uncertainties as black points. Each solution group is drawn in a different color and each peak of a distribution corresponds to an accidental ambiguity. In addition, the corresponding values from EtaMAID2018 \cite{Tiator:2018heh} (dashed line), BnGa-2019 \cite{muller2020new} (dotted line) and JüBo-2022 \cite{rnchen2022light} (dash-dotted line) are shown as well. \label{fig::reproduced_data_lmax_1}}
    \end{figure*}

    At first, the regression was conducted using $\ell_\text{max}=1$.
    For each of the six energy-bins, the number of warmup and post-warmup samples was set to $2 \cross 10^4$, respectively.
    In total, $N_\text{c}=10$ chains are started at each solution, found via the Monte Carlo maximum a posteriori approach.
    The corresponding MCMC convergence diagnostics, displayed in \cref{sec:mcmc_convergence_diagnostics} \cref{fig:diagnostics_lmax1}, support this decision, with $\hat{R} < 1.01$ and relative Monte Carlo standard error within few percent or less.
    \\\\
    For each energy-bin above $950$~MeV, specifically $1050, 1150$ and $1250$~MeV, the measured $\sigma_0$ data are systematically higher for $\cos(\theta) > 0$ compared to the TPWA predictions.
    Furthermore, the TPWA predictions for $\check{\Sigma}$ do not resemble the original data points at all, as shown in \cref{fig::reproduced_data_lmax_1}.
    It appears that the statistical model utilized with truncation order $\ell_\text{max}=1$ can not adequately replicate data points for all observables.
    An elucidation for this phenomenon is provided in \cref{sec::nstar}.
    \\\\
    To enhance the data description flexibility of the TPWA model, the truncation order was increased to $\ell_\text{max}=2$, and the regression was re-executed.
    To obtain the desired MCMC convergence diagnostics for each of the six energy-bins, it was necessary to increase the number of warm-up and post-warm-up samples to $5 \cross 10^4$, respectively.
    $N_\text{c}$ remains the same as for $\ell_\text{max}=1$.
    The corresponding MCMC convergence diagnostics are displayed in \cref{sec:mcmc_convergence_diagnostics}, \cref{fig:diagnostics_lmax2}.
    Special phenomena that occur are detailedly discussed in \cref{sec:mcmc_convergence_diagnostics}.
    The TPWA model with a truncation order of $\ell_\text{max}=2$ effectively describes the original data points as evidenced by the reproduced data distributions in \cref{fig::reproduced_data_lmax_2}.
    \\\\
    In general, it is preferable to set the truncation order $\ell_\text{max}$ as high as possible, because lower partial waves can interfere with higher ones, leading to non-negligible contributions.
    However, increasing the truncation order also increases the number of accidental ambiguities.
    For example, with $\ell_\text{max}=3$ and $1250$~MeV, $43$ posterior modes were identified.
    This results in a situation that demands a large number of numerical computations to achieve the desired MCMC convergence diagnostics.
    Additionally, the visual assessment of clustering becomes challenging due to the large number of required chains.
    Furthermore, the statistical quality of the combined data sets do not permit observations of any F-wave contributions, such as those from the $N(1680)5/2^+$ \cite{pdg} resonance at $E^\text{lab}_\gamma \approx 1035$~MeV.
    Due to these considerations, this paper focuses on $\ell_\text{max}=2$, while truncation orders with $\ell_\text{max}>2$ are reserved for future research.

    \subsection{Indication of $N^*$ resonances}\label{sec::nstar}
    To summarize, the model with $\ell_\text{max}=1$ is inadequate in replicating the original data points for $\sigma_0$ and $\check{\Sigma}$ for the energies above $950$~MeV.
    This phenomenon could be explained by an emerging resonance in the energy range between $950$ and $1050$~MeV that couples to an orbital angular momentum $\ell > 1$ and predominantly contributes to $\sigma_0$ and $\check{\Sigma}$.
    Since isospin is a conserved quantity in the strong interaction, the reaction of $\eta$-photoproduction serves as an isospin-filter, meaning that for the subsequent discussion, only $N^*$ resonances require consideration.
    There are two $N^*$ resonances which fulfill the conservation laws, couple to $\ell = 2$, and fall within the required energy range (taking into account the Breit-Wigner width \cite{pdg} of the resonances).
    These resonances are the $N(1675) 5/2^-$ \cite{pdg} at $E^\text{lab}_\gamma \approx 1026$~MeV and the $N(1700) 3/2^-$ \cite{pdg} at $E^\text{lab}_\gamma \approx 1071$~MeV.
    \\\\
    There is also a resonance which opens up already at $E^\text{lab}_\gamma \approx 762$~MeV, specifically $N(1520) 3/2^-$ \cite{pdg}.
    However, this resonance has a branching ratio to $\eta N$ \cite{pdg} that is $\sim 10$ times smaller than the ones of $N(1675) 5/2^-$ and $N(1700) 3/2^-$.
    The data sets employed do not appear to possess the necessary sensitivity to see a contribution of $N(1520) 3/2^-$.

    \subsection{Multipole parameters}
\begin{figure*}[p]
        \includegraphics[width=\textwidth,height=\textheight,keepaspectratio]{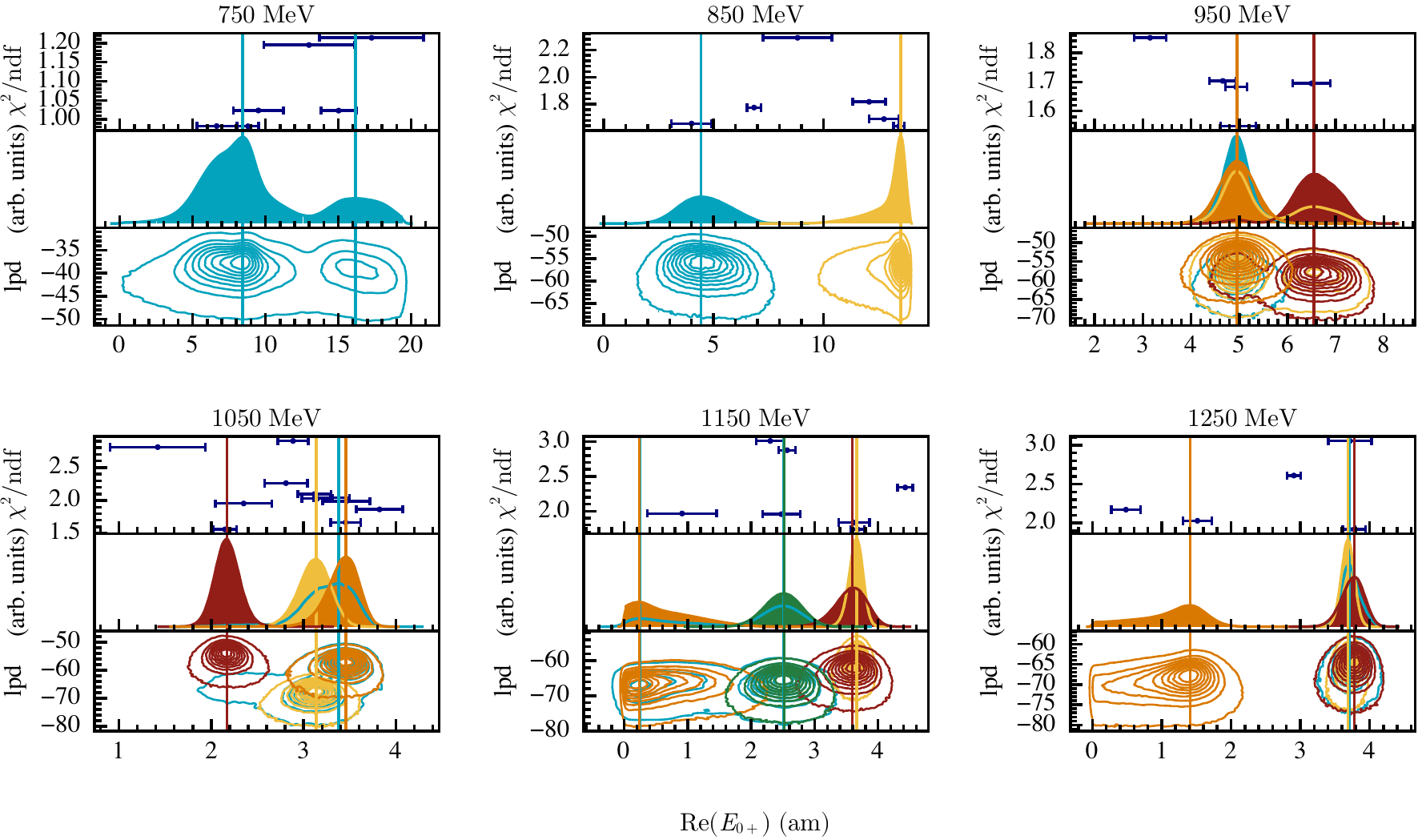}

        \vspace{10mm}

        \includegraphics[width=\textwidth,height=\textheight,keepaspectratio]{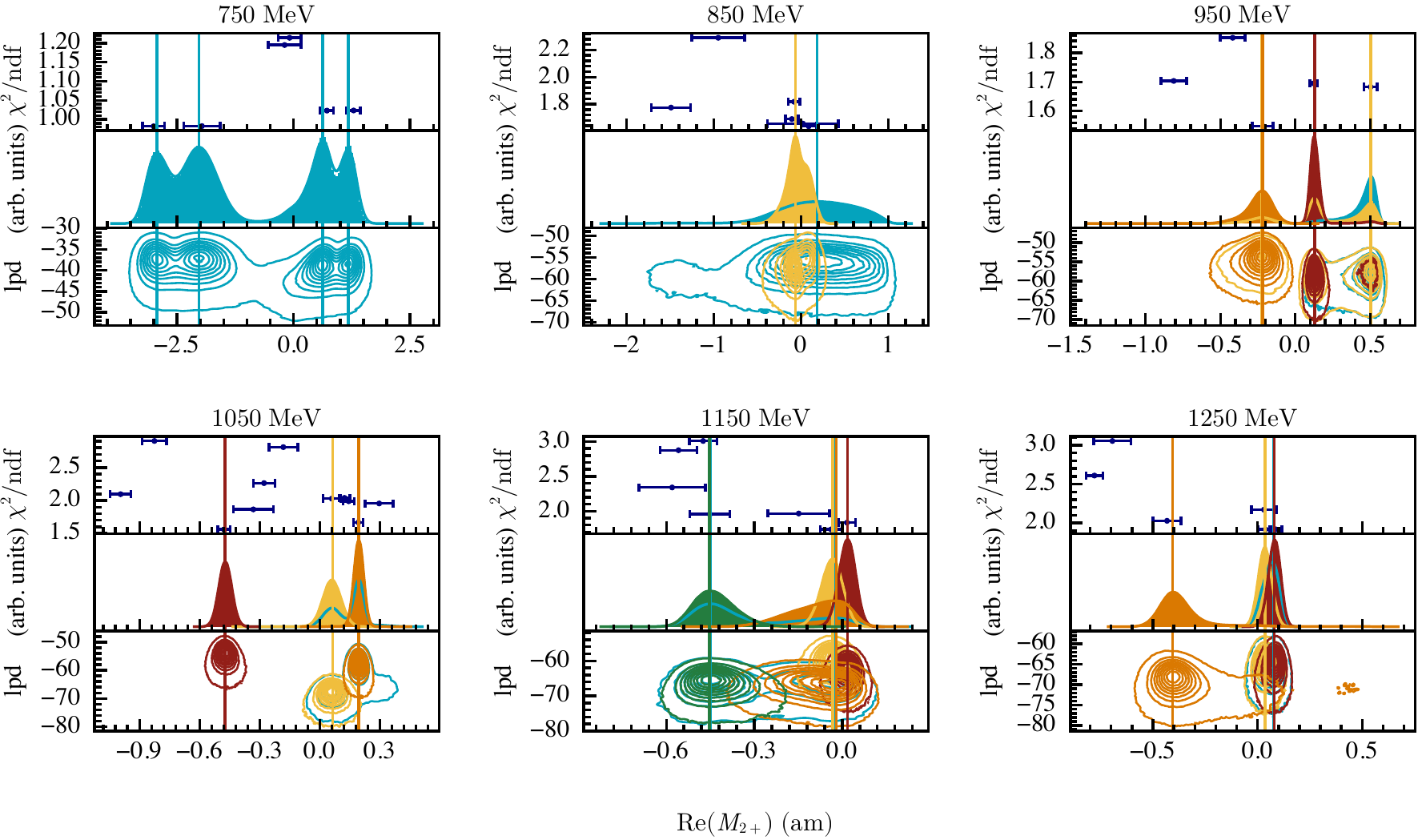}
        \caption{Solutions of the multipole parameters $\Re(E_{0+})$ and $\Re(M_{2+})$ for a truncation order of $\ell_\text{max}=2$, for the energy-bins $E^\text{lab}_\gamma=\qty[750,850,950,1050,1150,1250]$~MeV. Each solution group is drawn in a different color and each peak of a distribution corresponds to an accidental ambiguity. The different parts of the tripartite plots are explained at the beginning of \cref{sec:results}. The natural logarithm was used to calculate the log posterior density (lpd). \label{fig::multipole_plots_lmax_2_e0p}}
    \end{figure*}

    The solutions for $E_{0+}$ and $M_{2+}$ are shown as representative examples of the multipole parameters in \cref{fig::multipole_plots_lmax_2_e0p,fig::multipole_plots_lmax_2_m2p_2}.
    The figures for all multipole parameters are available in the supplementary material Ref.~\cite{supplementMaterial}.
    Typically, the peaks of the marginal distributions are in agreement with the first few 'best' a posteriori estimates.
    However, not every a posteriori solution has a corresponding peak within the marginal distributions.
    This could be due to two potential reasons.
    On the one hand, the interpretation of a marginal distribution differs from that of a maximum a posteriori estimate.
    On the other hand, the reason may lie within the Hamiltonian Monte Carlo algorithm \cite{duane1987hybrid, neal2011mcmc}, where it has been observed that some of the starting values are not in close proximity to the 'typical set'\footnote{The 'typical set' are the regions of the posterior which contribute the most to its expectation value.} \cite{betancourt2017}, but adjust rapidly.
    An example is shown in \cref{fig:converged_or_solution}.
    \\\\
    Within \cref{fig::all_energy_multipole_lmax_2}, the fifteen multipole parameters are graphed based on the laboratory energy of the photon.
    The corresponding values of EtaMAID2018 \cite{Tiator:2018heh}, BnGa-2019 \cite{muller2020new} and JüBo-2022 \cite{rnchen2022light} are also shown.
    For a detailed comparison of the various solution clusters and their relevance, readers are encouraged to refer to the tripartite multipole parameter figures in \cref{fig::multipole_plots_lmax_2_e0p,fig::multipole_plots_lmax_2_m2p_2} and the supplementary material in Ref.~\cite{supplementMaterial}.

    \subsubsection*{Comparison with MAID, BnGa and JuBo}
    In general, the paper's results align well with the values of EtaMAID2018, BnGa-2019 and JüBo-2022.
    However, there are two noteworthy exceptions.
    \\\\
    First, for the multipole parameter $\Im(E_{2+})$, none of the PWA values align with each other nor with the TPWA results of this paper.
    Second, the three PWAs report a value of $\sim 20$~mfm for the multipole parameter $\Re(E_{0+})$ at $750$~MeV.
    This significant value of $E_{0+}$ close to the $\eta p$-production threshold results from the dominant $N(1535)1/2^-$ resonance, which couples to the $S$-wave $E_{0+}$.
    The data sets of $\eta$-photoproduction (see \cref{sec:database}) utilized in this analysis do not emphasize such high values.
    Although the marginal parameter distribution does indeed have a non-zero probability at $\sim 19$~mfm, the most likely values are around $8.5$~mfm.
    This unexpected difference may have multiple causes.
    On the one hand, both BnGa-2019 and JüBo-2022 are coupled-channel analyses that involve a variety of final states simultaneously \cite{anisovich2016impact}.
    On the other hand, EtaMAID2018, BnGa-2019 and JüBo-2022 use the $\pi N$ partial-wave amplitudes from SAID \cite{anisovich2016impact} as input, which includes the $N(1535)1/2^-$ resonance \cite{PhysRevC.86.035202}.
    \\\\
    In contrast, the current analysis is not a coupled-channel analysis nor does it rely on the SAID solutions.
    The values of the multipole parameters are exclusively obtained from the $\eta$-photoproduction data sets presented in \cref{sec:database}.
    \\\\
    In addition, the TPWA relies on single-energy regression, which implies that solely the available data points at a particular energy-bin can be utilized in the analysis.
    As the PWAs do not follow this restriction, the complete available data sets can be utilized.
    This is particular important for the differential cross section, where the increased data can have a significant impact on the regression.
    \\\\
    Furthermore, although a dominant $S$-wave $E_{0+}$ results in a nearly constant maximal allowed value of one for the observable $E$ for all angles, the inverse conclusion does not always hold true.
    For example, the observable $E$ can not differentiate between $S$-wave $E_{0+}$ and $P$-wave $M_{1-}$ since both can lead to these maximum values of $E$.
    As shown in our results for $750$~MeV (see \cref{fig::all_energy_multipole_lmax_2}), the expected strength of $E_{0+}$ has migrated to other multipoles, such as $M_{1-}$.
    \\\\
    Improved statistics of the involved data sets in terms of the angular range or the inclusion of additional observables in future analyses, may shift the probability mass of the distribution of $\Re(E_{0+})$ at $750$~MeV towards the values of EtaMAID2018, BnGa-2019 and JüBo-2022.

    \subsection{Systematic parameters}
    The systematic parameters are all around the intended value of one.
    Each marginal distribution, for all systematic parameters, for all analyzed energy-bins, is exclusively unimodal.
    Examples can be found in \cref{fig::multipole_plots_lmax_2_m2p_2} and the supplementary material Ref.~\cite{supplementMaterial}.

    \subsection{Predicted data distributions}\label{sec::predicted_data}
\begin{figure*}[p]
        \includegraphics[width=\textwidth, height=\textheight, keepaspectratio]{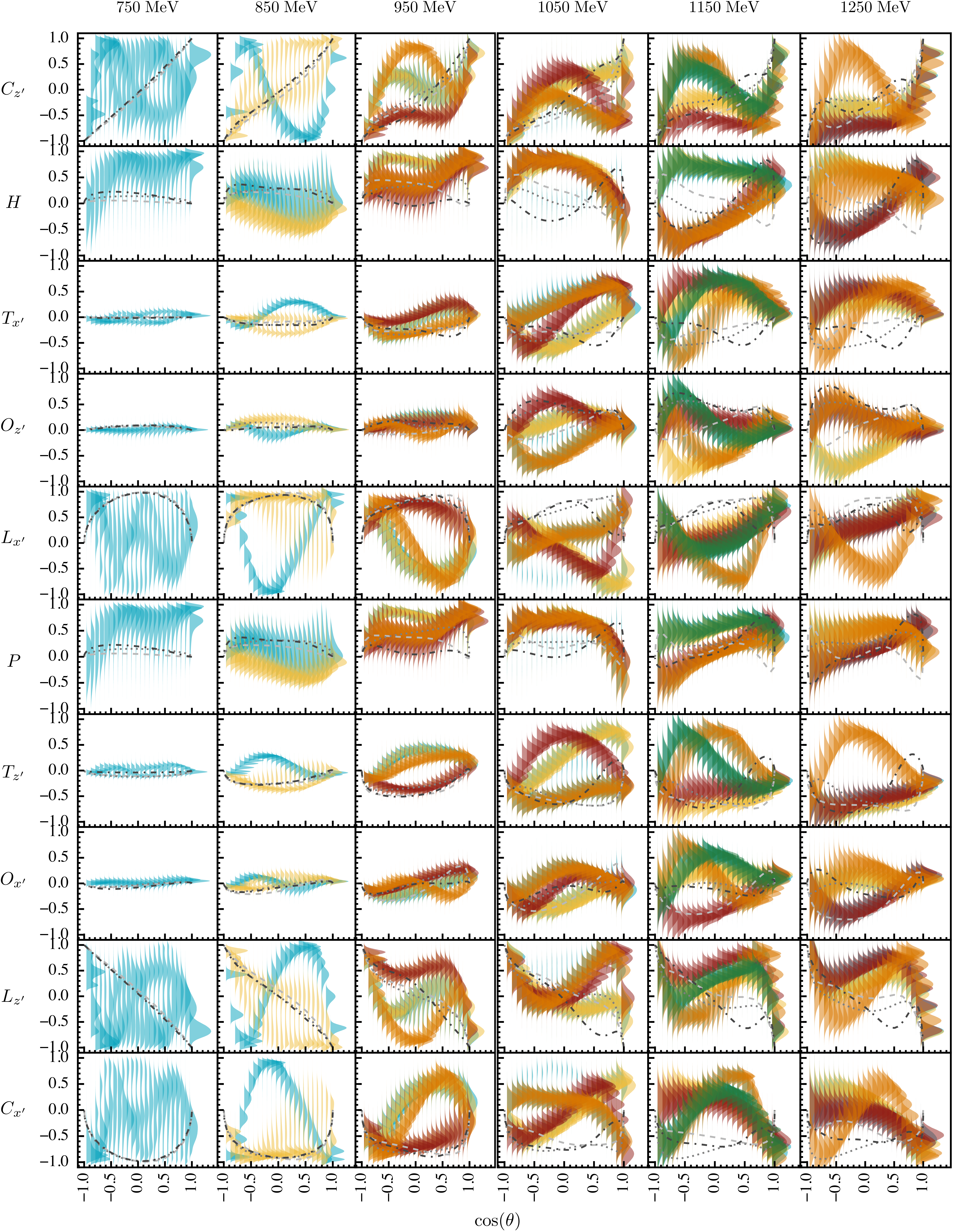}
        \caption{Predicted data distributions for the polarization observables $C_{z'}, H, T_{x'}, O_{z'}, L_{x'}, P, T_{z'}, O_{x'}, L_{z'}$ and $C_{x'}$ for the energy-bins $E^\text{lab}_\gamma=\qty[750,850,950,1050,1150,1250]$~MeV, using a truncation order of $\ell_\text{max} = 2$. Each solution group is drawn in a different color and each peak of a distribution corresponds to an accidental ambiguity. In addition, the corresponding values from EtaMAID2018 \cite{Tiator:2018heh} (dashed line), BnGa-2019 \cite{muller2020new} (dotted line) and JüBo-2022 \cite{rnchen2022light} (dash-dotted line) are shown as well. \label{fig::all_predicted_distributions_lmax_2}}
    \end{figure*}

    Using the estimated multipole parameters, predictions for polarization observables were calculated which were not utilized in this analysis.
    These include the observables $H$, $P$, as well as all eight observables of the groups $\mathcal{BR}$ and $\mathcal{TR}$ that have yet to be measured \cite{THIEL2022103949}.
    The predicted data distributions are displayed in \cref{fig::all_predicted_distributions_lmax_2}.
    The distributions are within the physical bounds between $-1$ and $1$ and their overall course over the angular range shows the correct tendency at $\cos(\theta) = \pm 1$ towards the mathematically expected values \cite{PhysRevC.89.055203}.
    \\\\
    An interesting effect can be observed.
    The predicted data distributions for the various ambiguities, show specific functional trajectories over the angular range.
    In contrast, for the reproduced data plots, the distributions were almost identical for different ambiguities.
    If there were experimental data related to one of the predicted observables that supports only one of the specific functional trends throughout the angular range, it would eliminate any other ambiguity.
    \\\\
    According to this criterion, potentially significant polarization observables have been chosen for upcoming experiments and are consolidated in \cref{tab::promising_candidates}.
    In particular, the polarization observable $C_{z'}$ seems suitable to reduce the ambiguities at all six energy-bins.
    \begin{table}
        \caption{Promising polarization observable candidates to resolve the ambiguities for truncation order $\ell_\text{max} = 2$.
        The corresponding predicted data distributions are shown in \cref{fig::all_predicted_distributions_lmax_2}.\label{tab::promising_candidates}}
        \begin{ruledtabular}
            \begin{tabular}{cl}
                $E^\text{lab}_{\gamma}$ / MeV & Observables\\
                \midrule
                750 & $C_{z'}, C_{x'}, L_{x'}, L_{z'}$ \\
                850 & $C_{z'}, C_{x'}, L_{x'}, L_{z'}, T_{x'}, T_{z'}$ \\
                950 & $C_{z'}, C_{x'}, L_{x'}, L_{z'}, T_{z'}$ \\
                1050 & $C_{z'}, C_{x'}, L_{x'}, O_{z'}, T_{z'}$ \\
                1150 & $C_{z'}, O_{x'}, T_{x'}, T_{z'}$ \\
                1250 & $C_{z'}$
            \end{tabular}
        \end{ruledtabular}
    \end{table}

    \section{Summary and Outlook}
    A TPWA was conducted for $\eta$-meson photoproduction off the proton near the production threshold.
    Model-independent estimates of electromagnetic multipole parameters were determined, allowing the first model-independent calculation of predictions of unmeasured polarization observables.
    Based on these results, promising future measurements were identified with the aim of minimizing remaining ambiguities.
    \\\\
    The data sets used in this study demonstrate clear D-wave contributions above $E^\text{lab}_\gamma=950$~MeV, but are not sensitive to F-wave or higher partial-wave contributions.
    \\\\
    For the first time, this study combined TPWA with Bayesian inference.
    The posterior distributions were highly multimodal, necessitating adaptations to monitor the MCMC convergence diagnostics.
    Despite its simplicity and use of less data, the TPWA approach maintains model-independence and achieves results consistent with the PWAs of MAID2018, BnGa-2019, and JüBo-2022.
    \\\\
    In general, resonances can be extracted from multipole parameters.
    However, for a precise extraction of the resonance parameters the current resolution of the combined data sets is not sufficient.
    In a subsequent study, the TPWA approach could be combined with the Laurent+Pietarinen parametrization for every multipole parameter \cite{PhysRevC.108.014615} in order to extract resonance parameters.
    In addition, the role of the prior distribution with regard to resolving the mathematical ambiguities could be investigated.

    \begin{acknowledgments}
        The authors would like to thank Prof. Dr. Sebastian Neubert, Prof. Dr. Carsten Urbach, and Dr. Jan Hartmann for several fruitful discussions.
        Furthermore, special thanks go to Prof. Dr. Reinhard Beck for his support.
    \end{acknowledgments}

    \begin{figure*}[p]
        \includegraphics[width=\textwidth,height=\textheight,keepaspectratio]{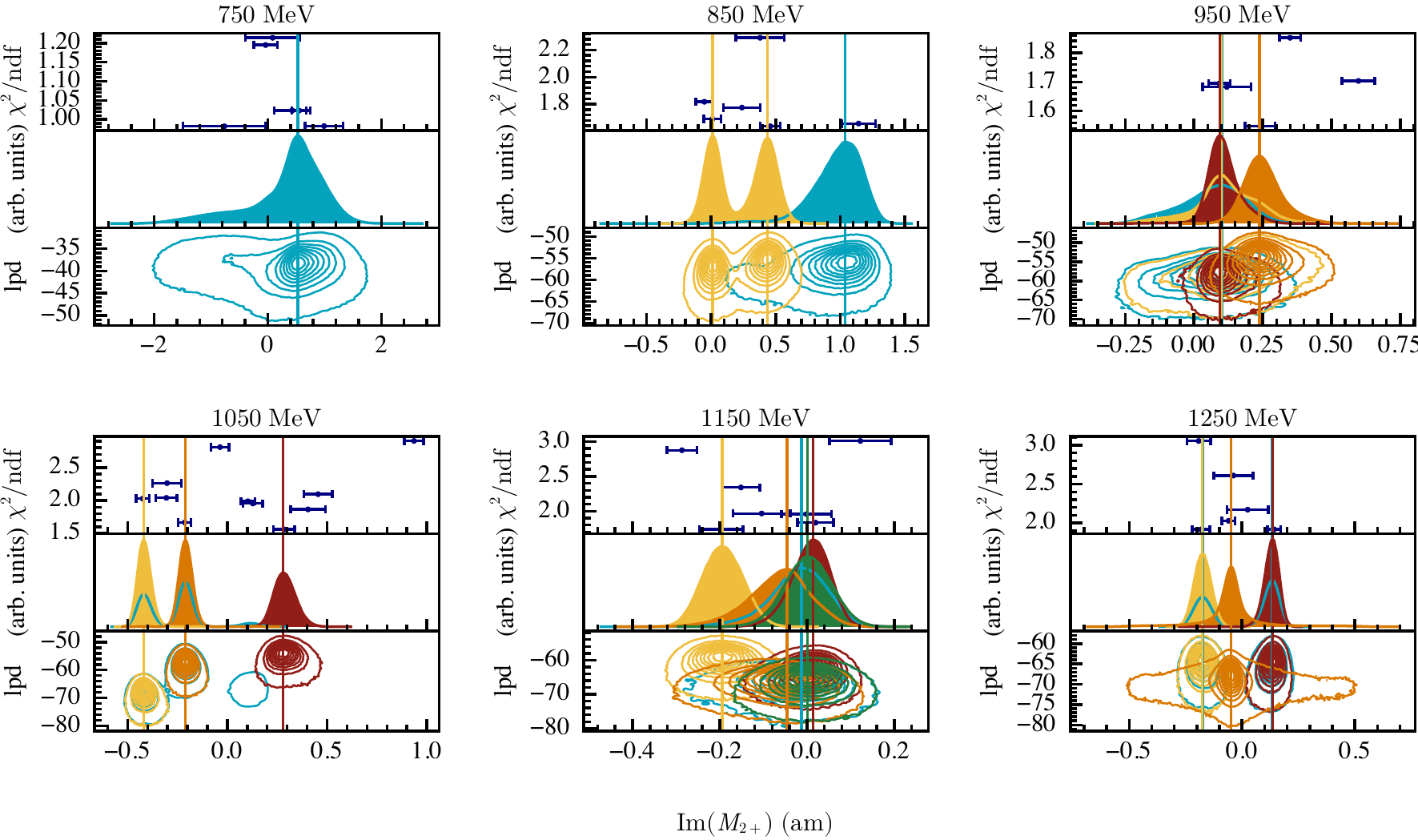}

        \vspace{10mm}

        \includegraphics[width=\textwidth,height=\textheight,keepaspectratio]{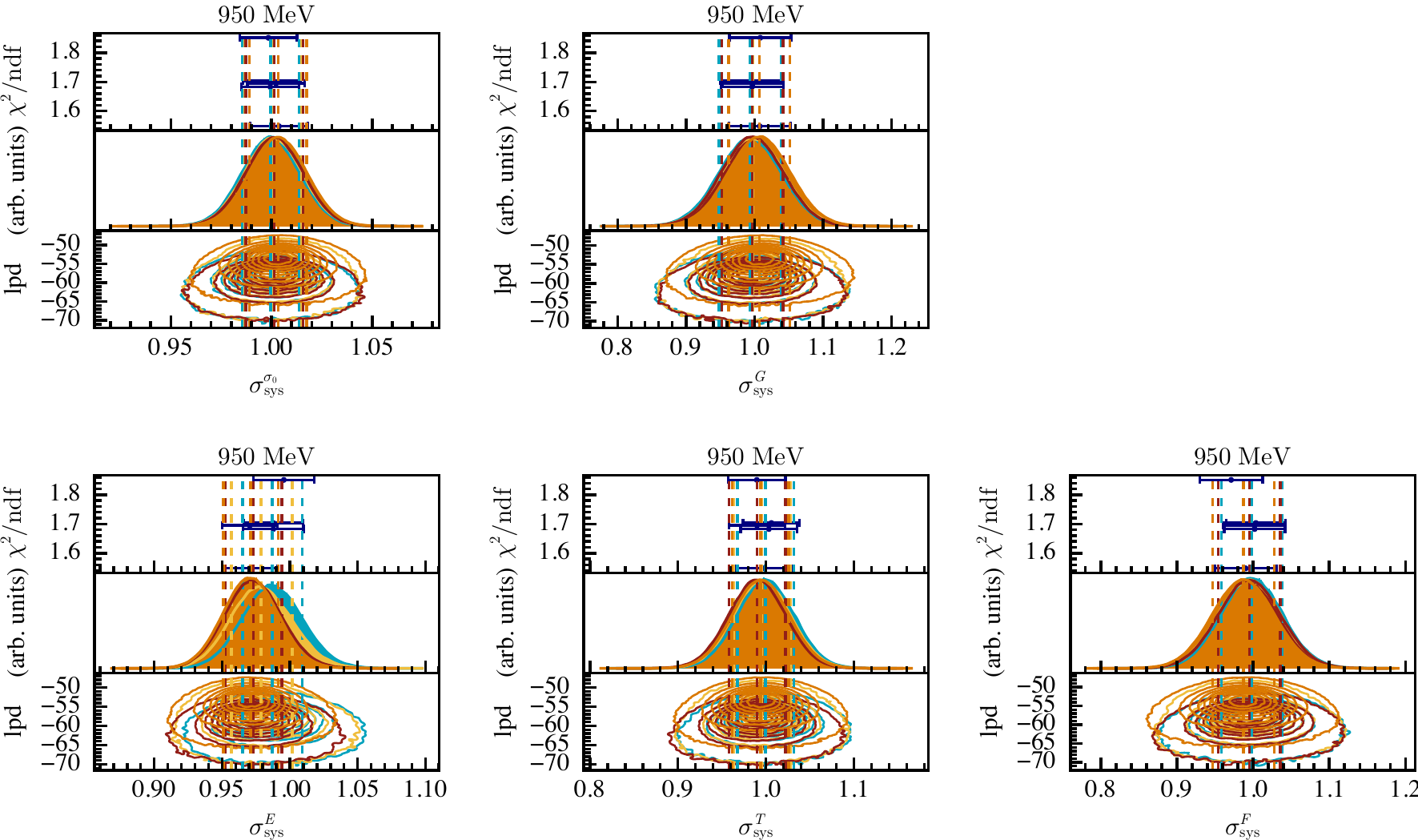}

        \caption{Solutions of the multipole parameter $\Im(M_{2+})$ for a truncation order of $\ell_\text{max}=2$, for the energy-bins $E^\text{lab}_\gamma=\qty[750,850,950,1050,1150,1250]$~MeV. In addition, solutions of systematic parameters for a truncation order of $\ell_\text{max}=2$, for the energy-bin $E^\text{lab}_\gamma=950$~MeV are shown. Each solution group is drawn in a different color and each peak of a distribution corresponds to an accidental ambiguity. The different parts of the tripartite plots are explained at the beginning of \cref{sec:results}. The natural logarithm was used to calculate the log posterior density (lpd). \label{fig::multipole_plots_lmax_2_m2p_2}}
    \end{figure*}
    \begin{figure*}[p]
        \includegraphics[width=\textwidth, height=0.95\textheight, keepaspectratio]{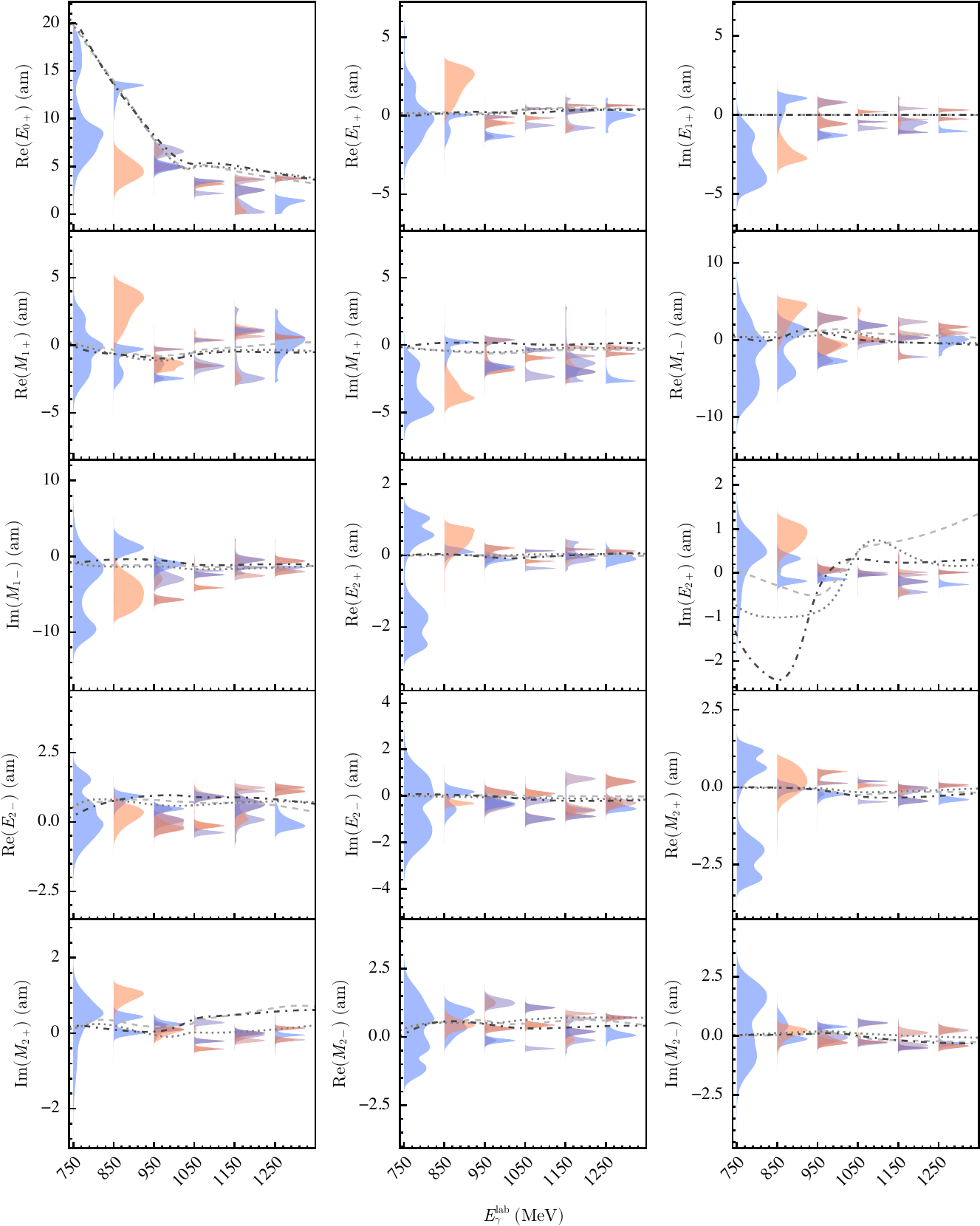}
        \caption{Marginal multipole solutions for the truncation order $\ell_\text{max}=2$ for the energy-bins $E^\text{lab}_\gamma=\qty[750,850,950,1050,1150,1250]$~MeV. In addition, the multipole parameter predictions from EtaMAID2018 \cite{Tiator:2018heh} (dashed line), BnGa-2019 \cite{muller2020new} (dotted line) and JüBo-2022 \cite{rnchen2022light} (dash-dotted line) are shown as well. The relevance of a solution is represented by a transition from sienna (less relevant) to blue (more relevant) hues. However, for a detailed comparison between the solutions and their relevance to each other, the reader is advised to the tripartite multipole parameter figures in \cref{fig::multipole_plots_lmax_2_e0p,fig::multipole_plots_lmax_2_m2p_2} and Ref.~\cite{supplementMaterial}. \label{fig::all_energy_multipole_lmax_2}}
    \end{figure*}

    \clearpage
    \appendix

    \section{Discrete ambiguities of the analyzed set of six polarization observables}\label{sec:DiscreteAmbSixObservables}
    Within this appendix, the discrete partial-wave ambiguities of the six observables $\qty{\sigma_{0}, \check{\Sigma}, \check{T}, \check{F}, \check{G}, \check{E}}$ analyzed within this work (cf. \cref{sec:database} and \cref{tab:used_data}) are discussed.
    It is argued that this specific set is mathematically complete in a TPWA.
    As has been demonstrated already in other works (e.g. Ref.~\cite{phdThesisYannick}), such mathematical considerations can still serve as a useful precursor to analyses of real data.

    The following discussion is based on the 'Omelaenko formalism' \cite{omelaenko1981ambiguities}.
    The basic definitions of the sixteen observables in pseudoscalar meson photoproduction, expressed in the transversity basis, are used.
    The expressions are collected in \cref{tab:observable_definitions}.
    \begin{table}[h]
        \caption{The definition of the sixteen polarization observables in terms of transversity amplitudes $b_{i}$ are displayed. The table is adapted from \cite{PhysRevC.55.2054}. The definition of the observables in terms of the required polarization configurations can be found in \cref{tab::polarization_configurations_observables}.
        \label{tab:observable_definitions}}
        \begin{ruledtabular}
            \begin{tabular}{lcc}
                Observable & Transversity-representation / $\rho$ & Type \\
                \midrule
                $ \check{\Omega}^{1} = \sigma_0 $ & $ \frac{1}{2} \left( \left| b_{1} \right|^{2} + \left| b_{2} \right|^{2} + \left| b_{3} \right|^{2} + \left| b_{4} \right|^{2} \right) $ & \\
                $ \check{\Omega}^{4} = - \check{\Sigma} $ & $ \frac{1}{2} \left( \left| b_{1} \right|^{2} + \left| b_{2} \right|^{2} - \left| b_{3} \right|^{2} - \left| b_{4} \right|^{2} \right) $ & $\mathcal{S}$ \\
                $ \check{\Omega}^{10} = - \check{T} $ &  $ \frac{1}{2} \left( - \left| b_{1} \right|^{2} + \left| b_{2} \right|^{2} + \left| b_{3} \right|^{2} - \left| b_{4} \right|^{2} \right) $ & \\
                $ \check{\Omega}^{12} = \check{P} $ &  $ \frac{1}{2} \left( - \left| b_{1} \right|^{2} + \left| b_{2} \right|^{2} - \left| b_{3} \right|^{2} + \left| b_{4} \right|^{2} \right) $ & \\
                & &  \\
                $ \check{\Omega}^{3} = \check{G} $ &  $ \mathrm{Im} \left[ - b_{1} b_{3}^{\ast} - b_{2} b_{4}^{\ast} \right] $ & \\
                $ \check{\Omega}^{5} = \check{H} $ &  $ \mathrm{Re} \left[ b_{1} b_{3}^{\ast} - b_{2} b_{4}^{\ast} \right] $ & $\mathcal{BT}$ \\
                $ \check{\Omega}^{9} = - \check{E} $ & $ \mathrm{Re} \left[ b_{1} b_{3}^{\ast} + b_{2} b_{4}^{\ast} \right] $ & \\
                $ \check{\Omega}^{11} = \check{F} $ &  $ \mathrm{Im} \left[ b_{1} b_{3}^{\ast} - b_{2} b_{4}^{\ast} \right] $ & \\
                & & \\
                $ \check{\Omega}^{14} = \check{O}_{x'} $ & $ \mathrm{Re} \left[ - b_{1} b_{4}^{\ast} + b_{2} b_{3}^{\ast} \right] $ & \\
                $ \check{\Omega}^{7} = - \check{O}_{z'} $ &  $ \mathrm{Im} \left[ - b_{1} b_{4}^{\ast} - b_{2} b_{3}^{\ast} \right] $  & $\mathcal{BR}$ \\
                $ \check{\Omega}^{16} = - \check{C}_{x'} $ & $ \mathrm{Im} \left[ b_{1} b_{4}^{\ast} - b_{2} b_{3}^{\ast} \right] $ & \\
                $ \check{\Omega}^{2} = - \check{C}_{z'} $ & $ \mathrm{Re} \left[ b_{1} b_{4}^{\ast} + b_{2} b_{3}^{\ast} \right] $ & \\
                & &  \\
                $ \check{\Omega}^{6} = - \check{T}_{x'} $ &  $ \mathrm{Re} \left[ - b_{1} b_{2}^{\ast} + b_{3} b_{4}^{\ast} \right] $ & \\
                $ \check{\Omega}^{13} = - \check{T}_{z'} $ &  $ \mathrm{Im} \left[ b_{1} b_{2}^{\ast} - b_{3} b_{4}^{\ast} \right] $ & $\mathcal{TR}$ \\
                $ \check{\Omega}^{8} = \check{L}_{x'} $ & $ \mathrm{Im} \left[ - b_{1} b_{2}^{\ast} - b_{3} b_{4}^{\ast} \right] $ & \\
                $ \check{\Omega}^{15} = \check{L}_{z'} $ & $ \mathrm{Re} \left[ - b_{1} b_{2}^{\ast} - b_{3} b_{4}^{\ast} \right] $ & \\
            \end{tabular}
        \end{ruledtabular}
    \end{table}

    \subsection{Discrete ambiguities of the group $\mathcal{S}$ observables in truncated partial-wave analysis}\label{sec:BasicRootDecomposition}
    As is well-known from Omelaenko's work, in the case of a truncated partial wave analysis with maximum angular momentum $\ell_{\text{max}}$, the four transversity amplitudes can be expressed in terms of linear factorizations:
    \begin{align}
        b_{1} \left(\theta\right) &= - \hspace*{1pt} \mathcal{C} \hspace*{1pt} a_{2L} \hspace*{1pt} \frac{\exp \left(- i \frac{\theta}{2}\right)}{\left( 1 + t^{2} \right)^{L}} \hspace*{1pt} \prod_{k = 1}^{2L} \left( t + \beta_{k} \right) \mathrm{,}  \\  \hspace*{5pt}
        b_{2} \left(\theta\right) &= - \hspace*{1pt} \mathcal{C} \hspace*{1pt} a_{2L} \hspace*{1pt} \frac{\exp \left(i \frac{\theta}{2}\right)}{\left( 1 + t^{2} \right)^{L}} \hspace*{1pt} \prod_{k = 1}^{2L} \left( t - \beta_{k} \right)  \mathrm{,}  \\
        b_{3} \left(\theta\right) &= \mathcal{C} \hspace*{1pt} a_{2L} \hspace*{1pt} \frac{\exp \left(- i \frac{\theta}{2}\right)}{\left( 1 + t^{2} \right)^{L}} \hspace*{1pt} \prod_{k = 1}^{2L} \left( t + \alpha_{k} \right) \mathrm{,} \hspace*{5pt}  \\
        b_{4} \left(\theta\right) &= \mathcal{C} \hspace*{1pt} a_{2L} \hspace*{1pt} \frac{\exp \left(i \frac{\theta}{2}\right)}{\left( 1 + t^{2} \right)^{L}} \hspace*{1pt} \prod_{k = 1}^{2L} \left( t - \alpha_{k} \right)  \mathrm{,}
    \end{align}
    where $t = \tan \frac{\theta}{2}$ (with the center-of-mass scattering angle $\theta$) and $\qty{\alpha_{k}, \beta_{k}}$ are the Gersten/Omelaenko-roots which are, in essence, equivalent to multipoles.

    Furthermore, all permissible solutions have to satisfy Omelaenko's constraint, i.e. \cref{eq:consistency}.
    The solution theory for the case where all four group $\mathcal{S}$ observables have been selected, and thus only ambiguities of the four moduli $\left| b_{1} \right|, \left| b_{2} \right|, \left| b_{3} \right|, \left| b_{4} \right|$ have to be considered, has been worked out at length in Ref.~\cite{phdThesisYannick}.
    This solution theory leads to the known complete sets of five (e.g.: $\qty{\sigma_{0}, \check{\Sigma}, \check{T}, \check{P}, \check{F}}$).
    In the following subsection, the special case where less than four diagonal observables are selected is considered.

    \subsection{Discrete ambiguities of the three group $\mathcal{S}$ observables $\qty{\sigma_{0}, \Sigma, T}$}\label{sec:AmbiguitiesOfThreeDiagonalObservables}
    The set of observables, used within this work, contains only three simultaneously diagonalized observables ($\sigma_0, \check{\Sigma}, \check{T}$, see \cref{tab:observable_definitions}).
    Therefore, one has to investigate which kinds of discrete ambiguities are allowed by this set of three observables, using the root-formalism described in \cref{sec:BasicRootDecomposition}.
    For this purpose, one can look at the 'minimal' linear combinations of squared moduli:
    \begin{align}
        \sigma_{0} - \check{\Sigma} &=  2 \left( \left| b_{1} \right|^{2} + \left| b_{2} \right|^{2} \right)  , \label{eq:Sigma0GEq1} \\
        \sigma_{0} + \check{\Sigma} &= 2 \left( \left| b_{3} \right|^{2} + \left| b_{4} \right|^{2} \right)  , \label{eq:Sigma0GEq2} \\
        \sigma_{0} + \check{T} &= 2 \left( \left| b_{1} \right|^{2} + \left| b_{4} \right|^{2} \right)  , \label{eq:Sigma0FEq1} \\
        \sigma_{0} - \check{T} &= 2 \left( \left| b_{2} \right|^{2} + \left| b_{3} \right|^{2} \right)  , \label{eq:Sigma0FEq2} \\
        - \check{\Sigma} + \check{T} &= 2 \left( \left| b_{1} \right|^{2} - \left| b_{3} \right|^{2} \right)  , \label{eq:GFEq1} \\
        - \check{\Sigma} - \check{T} &= 2 \left( \left| b_{2} \right|^{2} - \left| b_{4} \right|^{2} \right)  . \label{eq:GFEq2}
    \end{align}
    Upon reducing the problem to the non-redundant amplitudes $b_2$ and $b_4$ in the TPWA (by using $b_4 (W, \theta) = b_3 (W, - \theta)$ and $b_2 (W, \theta) = b_1 (W, -\theta)$, cf. \crefrange{eq:b1LinFactDecomp}{eq:b4LinFactDecomp}), one obtains:
    \begin{align}
        \sigma_{0} - \check{\Sigma} &\propto \prod_{k=1}^{2\ell_\text{max}} (t + \alpha^{\ast}_{k} ) (t + \alpha_{k} ) + \prod_{k=1}^{2\ell_\text{max}} (t - \alpha^{\ast}_{k} ) (t - \alpha_{k} ), \label{eq:b2b4Sigma0GEqRoots1} \\
        \sigma_{0} + \check{\Sigma} &\propto \prod_{k=1}^{2\ell_\text{max}} (t + \beta^{\ast}_{k} ) (t + \beta_{k} ) + \prod_{k=1}^{2\ell_\text{max}} (t - \beta^{\ast}_{k} ) (t - \beta_{k} ), \label{eq:b2b4Sigma0GEqRoots2} \\
        \sigma_{0} + \check{T} &\propto \prod_{k=1}^{2\ell_\text{max}} (t + \alpha^{\ast}_{k} ) (t + \alpha_{k} ) + \prod_{k=1}^{2\ell_\text{max}} (t - \beta^{\ast}_{k} ) (t - \beta_{k} ), \label{eq:b2b4Sigma0FEqRoots3} \\
        \sigma_{0} - \check{T} &\propto \prod_{k=1}^{2\ell_\text{max}} (t - \alpha^{\ast}_{k} ) (t - \alpha_{k} ) + \prod_{k=1}^{2\ell_\text{max}} (t + \beta^{\ast}_{k} ) (t + \beta_{k} ), \label{eq:b2b4Sigma0FEqRoots4} \\
        - \check{\Sigma} + \check{T} &\propto \prod_{k=1}^{2\ell_\text{max}} (t + \alpha^{\ast}_{k} ) (t + \alpha_{k} ) - \prod_{k=1}^{2\ell_\text{max}} (t + \beta^{\ast}_{k} ) (t + \beta_{k} ), \label{eq:b2b4GFEqRoots5} \\
        - \check{\Sigma} - \check{T} &\propto \prod_{k=1}^{2\ell_\text{max}} (t - \alpha^{\ast}_{k} ) (t - \alpha_{k} ) - \prod_{k=1}^{2\ell_\text{max}} (t - \beta^{\ast}_{k} ) (t - \beta_{k} ). \label{eq:b2b4GFEqRoots6}
    \end{align}
    The problem is now to find out which kinds of discrete ambiguity-transformations, when applied to the roots $\qty{\alpha_{k}, \beta_{k}}$, leave the full set of quantities \crefrange{eq:b2b4Sigma0GEqRoots1}{eq:b2b4GFEqRoots6} invariant, while also satisfying the multiplicative constraint \cref{eq:consistency}.
    The first set of transformations which comes to mind is given by the well-known double ambiguity:
    \begin{equation}
        \alpha_{k} \rightarrow \alpha_{k}^{\ast} \hspace*{5pt} \text{and} \hspace*{5pt} \beta_{k} \rightarrow \beta_{k}^{\ast} \hspace*{5pt} \forall k = 1,\ldots,2\ell_{max}. \label{eq:DoubleAmbTrafoSet2EqII}
    \end{equation}
    But other transformations may also be possible in addition, since the observable $\check{P}$ is missing from the full diagonalizable set $\qty{\sigma_{0}, \check{\Sigma}, \check{T}, \check{P}}$.
    Ideas that one would have to test are for instance exchange symmetries
    \begin{equation}
        \alpha_{k} \rightarrow \beta_{k} \hspace*{5pt} \text{and} \hspace*{5pt} \beta_{k} \rightarrow \alpha_{k} \hspace*{5pt} \forall k = 1,\ldots,2\ell_{max}, \label{eq:ExchangeSymmIdea}
    \end{equation}
    sign-changes
    \begin{equation}
        \alpha_{k} \rightarrow  - \alpha_{k} \hspace*{5pt} \text{and} \hspace*{5pt} \beta_{k} \rightarrow - \beta_{k} \hspace*{5pt} \forall k = 1,\ldots,2\ell_{max}, \label{eq:SignChangeIdea}
    \end{equation}
    or combinations of both
    \begin{equation}
        \alpha_{k} \rightarrow  - \alpha^{\ast}_{k} \hspace*{5pt} \text{and} \hspace*{5pt} \beta_{k} \rightarrow - \beta^{\ast}_{k} \hspace*{5pt} \forall k = 1,\ldots,2\ell_{max}. \label{eq:SignChangeIdeaII}
    \end{equation}
    All of these ideas indeed do not violate the constraint \cref{eq:consistency}.
    In case any such additional symmetry of the quantities \crefrange{eq:b2b4Sigma0GEqRoots1}{eq:b2b4GFEqRoots6} were found, the next step would be to test which of the remaining three observables $\qty{F, G, E}$ resolves the symmetry.
    Neither of the proposed symmetries \cref{eq:ExchangeSymmIdea,eq:SignChangeIdea,eq:SignChangeIdeaII} leaves all the six quantities \crefrange{eq:b2b4Sigma0GEqRoots1}{eq:b2b4GFEqRoots6} invariant.
    It remains to be asked whether such additional symmetries actually exist.
    In case they do not exist, the discussion would be simplified significantly (since $\check{F}$ and $\check{G}$ in this case already resolve the double ambiguity \cref{eq:DoubleAmbTrafoSet2EqII}).
    Due to information-theoretical reasons, it only seems permissible to simultaneously use three of the quantities from \crefrange{eq:b2b4Sigma0GEqRoots1}{eq:b2b4GFEqRoots6}, i.e. to use three new quantities obtained via invertible and linear transformations from the three diagonal, initial observables $\qty{\sigma_{0}, \check{\Sigma}, \check{T}}$.

    As an example, one can select the three quantities given by \cref{eq:b2b4Sigma0GEqRoots1,eq:b2b4Sigma0GEqRoots2,eq:b2b4Sigma0FEqRoots3}.
    The full set of discrete ambiguity-transformations, which, when applied to the roots $\qty{\alpha_{k}, \beta_{k}}$, leaves \cref{eq:b2b4Sigma0GEqRoots1,eq:b2b4Sigma0GEqRoots2} invariant while maintaining the constraint in \cref{eq:consistency}, is given by the two transformations in \cref{eq:DoubleAmbTrafoSet2EqII,eq:SignChangeIdea}.
    Under the exchange symmetry \cref{eq:ExchangeSymmIdea}, \cref{eq:b2b4Sigma0GEqRoots1,eq:b2b4Sigma0GEqRoots2} are transformed into each other and thus are not invariant.

    Now considering additionally the quantity in \cref{eq:b2b4Sigma0FEqRoots3}, one can see that while the transformation \cref{eq:DoubleAmbTrafoSet2EqII} leaves this quantity invariant, transformation \cref{eq:SignChangeIdea} does not.
    This only leaves one possible conclusion, namely that also for the case of only three diagonal observables $\qty{\sigma_{0}, \check{\Sigma}, \check{T}}$, or equivalently the three new quantities in \cref{eq:b2b4Sigma0GEqRoots1,eq:b2b4Sigma0GEqRoots2,eq:b2b4Sigma0FEqRoots3}, the double ambiguity is the only relevant discrete ambiguity of the problem\footnote{This statement is of course only true in case transformations \cref{eq:DoubleAmbTrafoSet2EqII} and \cref{eq:SignChangeIdea} are indeed the only possible discrete ambiguities of the quantities in \cref{eq:b2b4Sigma0GEqRoots1,eq:b2b4Sigma0GEqRoots2} and that no further such discrete ambiguities exist. This seems plausible when considering equations \cref{eq:b2b4Sigma0GEqRoots1,eq:b2b4Sigma0GEqRoots2}, in combination with the constraint in \cref{eq:consistency}.}.

    The argument given above can be repeated for any other case where a combination of three quantities from the six definitions \crefrange{eq:b2b4Sigma0GEqRoots1}{eq:b2b4GFEqRoots6} is taken as a starting point.
    None of the other starting-combinations is necessary for a proof, since this would give a redundant derivation, with the same outcome.

    \subsection{Completeness of the set $\qty{\sigma_{0}, \check{\Sigma}, \check{T}, \check{F}, \check{G}, \check{E}}$}
    It has already been shown in Refs. \cite{phdThesisYannick,PhysRevC.89.055203} that the observables $\check{F}$ and $\check{G}$ change sign under the double-ambiguity transformation.

    All the arguments made up to this point prove that the set $\qty{\sigma_{0}, \check{\Sigma}, \check{T}, \check{F}, \check{G}, \check{E}}$ is free of discrete ambiguities in the TPWA.
    Assuming furthermore that this set of six observables has no continuous ambiguities, the set is complete.

    \clearpage
    \section{Covered phase space of the used data}\label{sec:data_coverage}
\begin{figure*}
        \centering
        \includegraphics[width=\textwidth]{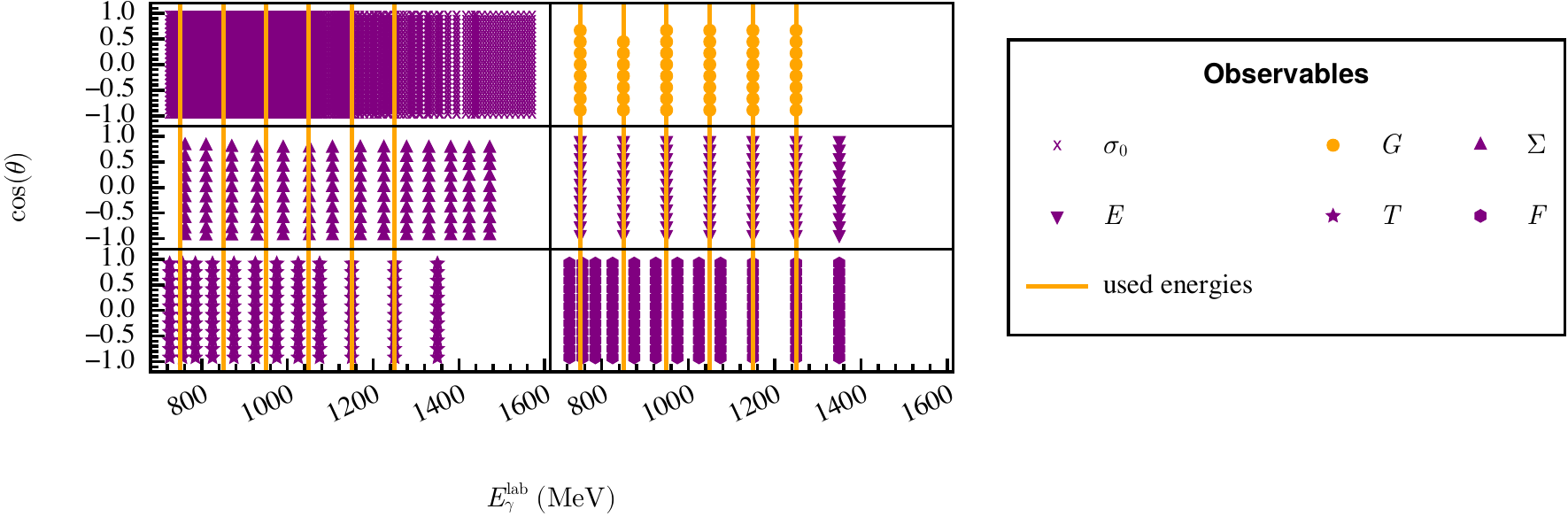}
        \caption{Energy and angular coverage of the six observables $\sigma_0, \Sigma, G, E , T$ and $F$ \cite{PhysRevLett.118.212001,bartalini2007measurement,muller2020new,afzal2019measurement,akondi2014measurement} which were used for the analysis. The used energies $E^\text{lab}_\gamma=[750,850,950,1050,1150,1250]$~MeV are determined by the observable $G$. \label{fig:energy_angular_coverage}}
    \end{figure*}

    The phase-space coverages of the used polarization observable data are illustrated in \cref{fig:energy_angular_coverage}.
    For a detailed description of the data see \cref{sec:database,tab:used_data}.
    The vertical orange lines correspond to the energy-bins of the statistically weakest polarization observable $G$ and indicate by which amount the data set of an other observable has to be shifted to match these energies.

    \section{On the correlation of profile functions}\label{sec:definitionCorrelation}
    The correlation of two random variables $X$ and $Y$ can be calculated using the Pearson correlation coefficient defined as \cite{definition_correlation}:
    \begin{equation}
        \mathrm{Corr}(X,Y) = \frac{\mathrm{Cov}(X,Y)}{\sqrt{\mathrm{Var}[X] \mathrm{Var}[Y]}}, \label{eq:correlation0}
    \end{equation}
    with their respective variances $\mathrm{Var}$ and the covariance $\mathrm{Cov}$ between the two random variables.
    Under the assumption that the dimensionless observables do not have any correlation with each other, the covariance of the unpolarized differential cross-section $\sigma_0(W,\theta)$ (denoted with $X$) and a profile function (denoted as $Y'=XY$, as $\sigma_0(W,\theta)$ was used to calculate the profile function) is:
    \begin{align}
        \mathrm{Cov}(X,Y') &= \mathrm{E}[XXY] - \mathrm{E}[X] \cdot \mathrm{E}[XY],\notag\\
        &= \qty (\mathrm{E}[X^2] - \mathrm{E}[X]^2 ) \cdot \mathrm{E}[Y],\notag\\
        &= \mathrm{Var}[X] \cdot \mathrm{E}[Y].\label{eq:covariance1}
    \end{align}
    And similarly for the covariance of one profile function (denoted as $Y'=XY$) to another (denoted as $Z'=XZ$):
    \begin{align}
        \mathrm{Cov}(Y',Z') &= \mathrm{E}[XYXZ] - \mathrm{E}[XY] \cdot \mathrm{E}[XZ],\notag\\
        &= \qty (\mathrm{E}[X^2] - \mathrm{E}[X]^2 ) \cdot \mathrm{E}[Y] \cdot \mathrm{E}[Z],\notag\\
        &= \mathrm{Var}[X] \cdot \mathrm{E}[Y] \cdot \mathrm{E}[Z].\label{eq:covariance2}
    \end{align}
    Substituting \cref{eq:covariance1}, and \cref{eq:covariance2} respectively, into \cref{eq:correlation0} the correlation for both cases is:
    \begin{align}
        \mathrm{Corr}(X,Y')  &= \sqrt{\frac{\mathrm{Var}[X]}{\mathrm{Var}[Y']}} \cdot \mathrm{E}[Y], \label{eq:correlation1}\\
        \mathrm{Corr}(Y',Z') &= \frac{\mathrm{Var}[X]}{\sqrt{\mathrm{Var}[Y'] \cdot \mathrm{Var}[Z']}} \cdot \mathrm{E}[Y] \cdot \mathrm{E}[Z]. \label{eq:correlation2}
    \end{align}

    \section{Underlying assumptions}\label{sec:assumptions}
    An enormous strength of Bayesian statistics is its clarity about the underlying assumptions and how these evolve into the used statistical model.
    In general one has $N$ data-pairs $\qty(y,x)_i$, where the two components can be distinguished as follows:
    \begin{enumerate}
        \item
        The random variables $\vb*{y} = \qty(y_1, \dots, y_N)$ follow a certain distribution.
        In this context, these correspond to the values of the profile functions of the polarization observables $\check{\Omega}^{\alpha}(W, \theta)$.
        \item
        The explanatory variables \cite{gelman2013bayesian} $\vb*{x} = \qty(x_1, \dots, x_N)$ do not belong to any probability distribution.
        In this context, these are the angular values $\cos(\theta_i)$ at which the $y_i$ were measured.
    \end{enumerate}
    The underlying distribution of $\vb*{y}$ is of upmost importance as it defines the shape of the likelihood function and, by association, the structure of the parameter space.
    It is therefore essential to examine the distribution from which $\vb*{y}$ originates and discuss the validity of the involved assumptions.
    Hereby, an understanding of the data-taking as well as the subsequent analysis, to extract values for the polarization observables, is mandatory.
    For this reason, special emphasis is placed on their discussion within this paper.

    The polarization observables used within this analysis, originate from measurements at multiple experimental facilities: ELSA \cite{hillert2006bonn}, MAMI \cite{kaiser20081} and GRAAL \cite{bartalini2005measurement}.
    The measured quantities are count rates, corresponding to differential cross-sections, from which then, one or multiple polarization observables can be extracted.
    The two most common methods are a 'binned chi-square fit' and an 'unbinned maximum-likelihood fit' \cite{afzal2019measurement}.
    For the first case, it is common to use an asymmetry of the form:
    \begin{equation}
        A \propto \frac{N_1 - N_2}{N_1 + N_2} \label{eq:asymmetry},
    \end{equation}
    where $N_1, N_2$ are normalized count rates of reconstructed $\gamma p \rightarrow \eta p$ events for different polarization states \cite{akondi2014measurement,bartalini2007measurement}.
    This has the advantage that systematic effects for example from the reconstruction efficiency cancel out.

    Certainly, the distribution of this asymmetry is not explicitly addressed in any of the analyses, concerning polarization observables, which the authors have encountered up to this point.
    However, since the distribution of $A$ determines the structure of the likelihood distribution, it is mandatory to study its proper form.

    The count rates $N_1, N_2$ are Poisson-distributed random variables.
    If the expectation value, typically denoted as $\lambda$, is high enough, the distribution goes over to a Gaussian distribution.
    In the case of the here used data, this should be a good assumption.

    The sum or difference of two independent Gaussian distributed random variables, as present in \cref{eq:asymmetry} is again Gaussian distributed, which can be shown for example using characteristic functions.

    However, the ratio of two, eventually correlated, Gaussian distributions $Z=X/Y$ is far more complicated.
    A general treatment can be found in Ref.~\cite{marsaglia2006ratios}.
    Additionally a closed form expression is given in \cref{eq:QuotientIntClosedExpressionMathematica}, \cref{sec:QuotientDistribution}.
    Indeed, there exist Gaussian shapes for the asymmetry $A$ in certain limits, but there exists also the possibility for a bimodal distribution \cite{marsaglia2006ratios}.
    Therefore, the shape of the asymmetry $A$ has to be checked for the absence of a bimodal structure.
    In order to use $\chi^2$ as likelihood function, the distribution should be well approximated by a Gaussian distribution.
    These checks can be performed by inserting the corresponding values for the expectation values ($\mu_x,\mu_y$), standard deviations ($\sigma_x, \sigma_y$) and correlation ($\rho$) into the formula for $Z$ and its transformation, see Ref.~\cite{marsaglia2006ratios}, or by using \cref{eq:QuotientIntClosedExpressionMathematica}.

    An alternative approach, where the utilization of such an asymmetry can be circumvented, is the already mentioned 'unbinned maximum-likelihood fit'.
    Albeit, in contrast to the first method, the detector acceptance has to be taken into account \cite{afzal2019measurement}, which is possible \cite{handle:20.500.11811/7258}.
    Within this approach, the likelihood distribution can be modeled appropriately using Poisson distributions.

    Summarizing, it is advantageous to use the 'unbinned maximum-likelihood fit' for future analyses, in order to extract values for the polarization observables.

    However, the distribution of the extracted polarization observables not only depends on the shape of the used likelihood function, but also implicitly on the method used to estimate the parameter uncertainties.
    Again, the distribution of the parameters is rarely explicitly discussed within papers such as the references cited in \cref{tab:used_data}.
    The error analysis of MINUIT uses by default the HESSE approach \cite{james1975minuit}, which assumes an asymptotic approximation to a Gaussian distribution for the parameters under consideration.
    Thus, it is likely that the parameters were assumed to be Gaussian distributed.
    Another indication in the same direction is that all data used within the present analysis (cf. \cref{tab:used_data}) do have symmetric statistical uncertainties \cite{PhysRevLett.118.212001,muller2020new,afzal2019measurement,akondi2014measurement,bartalini2007measurement}.

    The profile functions $\check{\Omega}^{\alpha}$ are calculated by a product of random variables.
    However, even when these two random variables are independent and Gaussian distributed, the result is not always a Gaussian, only when one of the standard deviations is very small, see \cite{phdThesisYannick} or \cref{sec:ProductDistribution}.
    Fortunately, this is the case for $\sigma_0$ as it is the observable in $p \eta$-photoproduction with an unprecedented accuracy.

    \section{Analysis steps}\label{sec:analysis_steps}
\begin{figure}[t]
        \includegraphics[width=\linewidth]{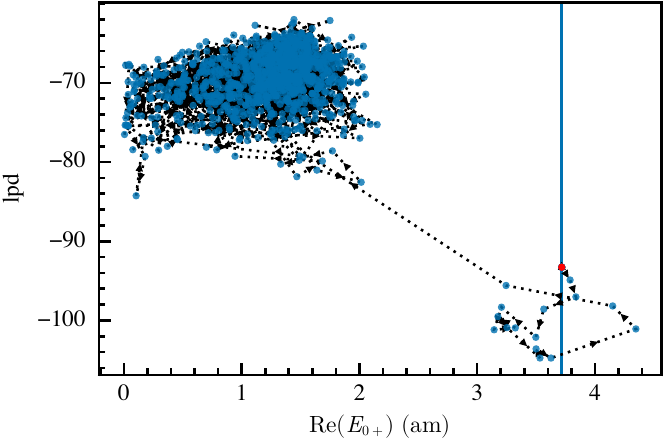}
        \caption{Illustration of the first 1000 sampling points of a chain with initial value at $3.7$ (the blue vertical line).
        The first sampling point is drawn in red.
        The chain converges from its starting point to a more likely solution, i.e. with higher log posterior density (lpd) value.
        The natural logarithm was used to calculate the lpd.
        \label{fig:converged_or_solution}
        }
    \end{figure}

    This section explains in detail the analysis steps in order to determine the complex multipole parameters using Bayesian inference.
    \\\\
    The posterior, which was in all of the analyses explicitly multimodal, and the goal to analyze the structure of the mathematical ambiguities, bear a major challenge with respect to the sampling of the posterior distribution.
    On the one hand, posteriors with multiple modes connected by regions of low posterior density persuade the Markov chains to get stuck within a certain mode, unable to explore multiple ones \cite{gelman2013bayesian}.
    This results in drastically\footnote{This behavior was to be expected since $\hat{R}$ is a measure whether all chains have converged to the same distribution.} failing MCMC convergence diagnostics, such as the potential-scale-reduction statistic $\hat{R}$.
    On the other hand, the number of possible modes increases exponentially with the truncation order $\ell_\text{max}$.
    An upper limit can be given by $2^{4 \ell_\text{max}} - 2$, as this is the maximal possible number of accidental ambiguities of the four group $\mathcal{S}$ observables (note that the bulk of this number is probably not realized as actual ambiguities, due to the multiplicative constraint \cref{eq:consistency}) \cite{phdThesisYannick}.
    Capturing consistently all modes of the marginal posterior distributions via a large number of chains, with randomized starting values is computationally inefficient.
    Furthermore, randomized starting values will lead to traceplots where one can not distinguish between chains that have not converged yet and chains which have explored more than one mode.
    An illustrative example is shown in \cref{fig:converged_or_solution}.
    These difficulties can be overcome by specifying well chosen starting values for the MCMC algorithm, explained in more detail in \cref{sec:posteriorSampling,sec:mcmcConvergence}.
    On that account, certain parts of the typical Bayesian workflow \cite{gelman2020bayesianWorkflow} had to be adapted.

    \subsection{Monte Carlo maximum a posteriori estimation}\label{section:mcmle}
    In order to compare between different solutions, found within the same analysis, it is important to find all modes of the marginal posterior distributions, especially the global maximum.
    As already mentioned, the number of accidental ambiguities rises exponentially with the truncation order.
    Thus, the utilization of an optimization routine is substantially more efficient\footnote{Integration is far more computation-intensive than differentiation.} than a large number of MCMC chains.
    With this in mind, a Monte Carlo maximum a posteriori estimation of the proposed posterior is employed as a preparatory step for the Bayesian inference procedure.
    The results of the following approach are cross checked via an implementation in Mathematica \cite{Mathematica}, using the Levenberg-Marquard-algorithm \cite{levenberg1944method,marquardt1963algorithm}, as well as in Julia \cite{bezanson2017julia}, using the L-BFGS-algorithm \cite{broyden1970convergence,fletcher1970new,goldfarb1970family,shanno1970conditioning,nocedal1980updating} via Optim.jl \cite{mogensen2018optim}.
    \\\\
    At first, one needs to fix the overall phase of the multipoles, due to the bilinear product in \cref{eq:bilinear_product}.
    Indeed, without such a constraint the minimization algorithms would have convergence problems, as the solutions are no longer located at isolated points in the parameter space but on continuous connected regions.
    Without loss of generality, a valid choice is $\Re(E_{0+}) > 0$, $\Im(E_{0+}) = 0$ \cite{phdThesisYannick}.
    Second, the minimization algorithm is performed for $n$ different starting values.
    The starting values are chosen within the physically allowed parameter space, which solely depends on the total cross-section $\sigma_\text{tot}$ \cite{phdThesisYannick,dreschsel1992threshold}.
    Fortunately, the unpolarized differential cross-section is the most accurately measured observable in $p\eta$-photoproduction \cite{THIEL2022103949}, thus yielding accurate limits.
    An appropriate amount of $n$ equidistant points is chosen on each axis of this $8 \ell_\text{max} - 1$ dimensional hyper-rectangle, such that the volume is sufficiently covered.
    Each of these parameter configurations is then used as starting values for the minimization algorithm.
    \\\\
    Finally, the non redundant solutions, of the $n$ possible mode candidates, can be extracted via a clustering algorithm.
    Hereby, all values of the multipole parameters are rounded to six digits.
    Then the unique solution vectors can be filtered out.
    A rough estimate for the uncertainty of each parameter solution is calculated via the inverse of the Hesse matrix \cite{barlow1993statistics}, i.e. assuming a Gaussian shape of the parameters.

    \subsection{Sampling of the posterior}\label{sec:posteriorSampling}
    Within this work, the well established probabilistic programming software Stan \cite{stan2022stan} has been used to encode the employed model and to run the posterior sampling with the state-of-the-art Hamiltonian Monte Carlo algorithm \cite{duane1987hybrid,neal2011mcmc} in combination with the No-U-Turn sampler \cite{hoffman2014no}.
    The employed Stan model can be found in the supplementary material Ref.~\cite{supplementMaterial}.
    \\\\
    For each mode of the posterior distribution, determined within \cref{section:mcmle}, $N_\text{c}$ chains are sampled with starting values for the multipole and systematic parameters equal to the corresponding $(8 \ell_\text{max} + 4)$-dimensional solution vector.
    This approach ensures adequate sampling of all marginal posterior modes and enables again a meaningful convergence diagnostics, further discussed in \cref{sec:mcmcConvergence}.
    Hence, this is true as long as the posterior modes are in the vicinity of the 'typical set' \footnote{An illustration of the 'typical set' can be found in \cite{betancourt2017}.}, which is the case within this paper.
    \\\\
    The following tuning-parameters of the Hamiltonian Monte Carlo algorithm and the No-U-Turn sampler are adapted to the problem at hand.
    The average Metropolis acceptance probability $\delta \in [0,1]$ is increased from its default value of 0.8 to $\delta = 0.99$.
    Thus, preferring a more fine-grained sampling, i.e. smaller leapfrog\footnote{This refers to one parameter of the leapfrog integrator, see for example \cite{neal2011mcmc}.} steps $\epsilon$ \cite{hoffman2014no}, over the additional computation time.
    The maximum tree depth, with a default value of 10, is increased to 50, so that the algorithm can explore even challenging posterior regions without hitting the termination conditions \cite{stan2022stan}.

    \subsection{Monitor MCMC convergence}\label{sec:mcmcConvergence}
\begin{figure}
        \centering
        \includegraphics[width=\linewidth]{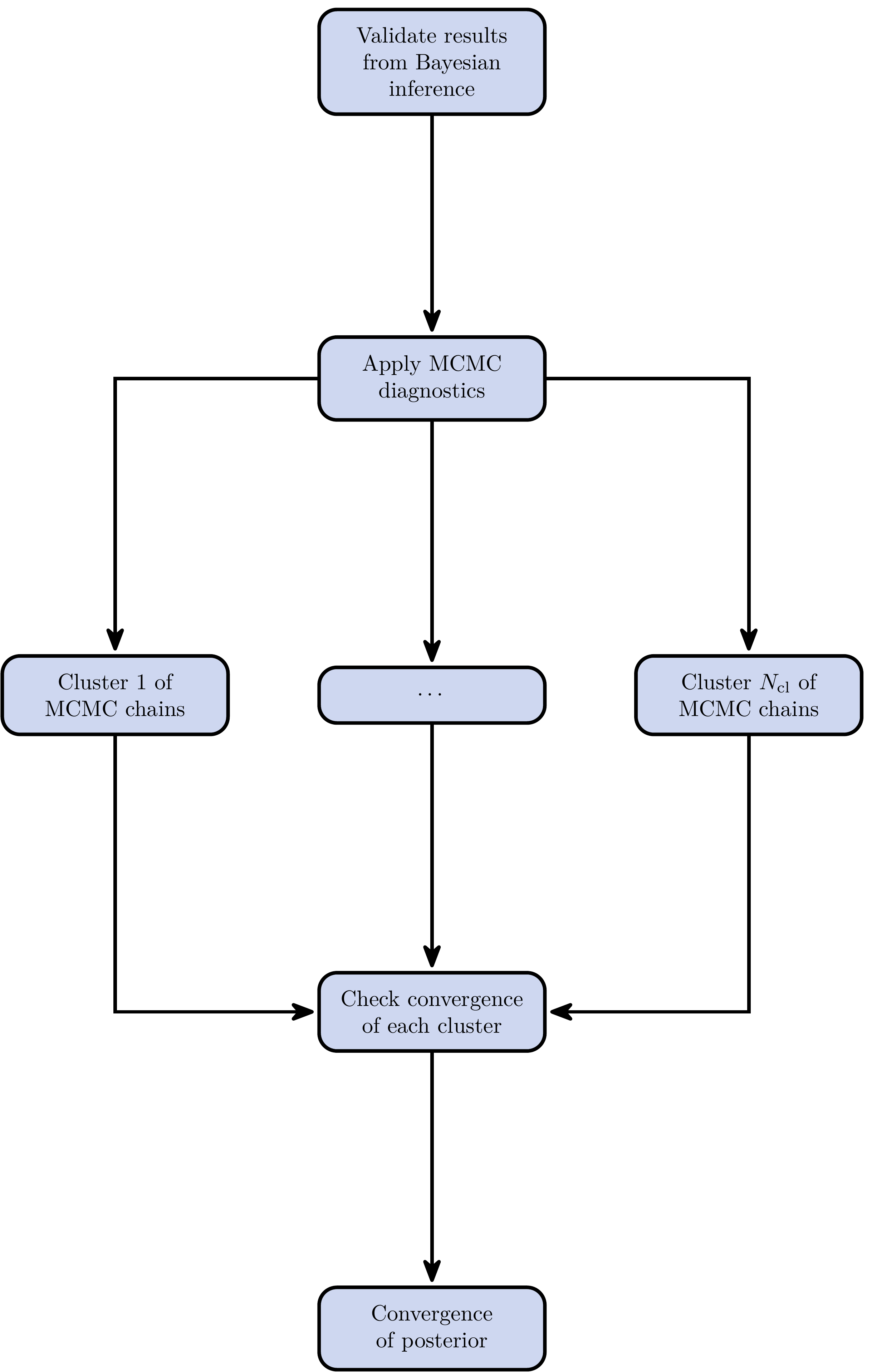}
        \caption{Adapted workflow to monitor MCMC convergence due to a multimodal posterior. \label{fig:schematic_mcmc_convergence}}
    \end{figure}

    Naturally one is interested in how well the structure of the posterior was explored by the applied MCMC algorithm.
    The goal is to diagnose whether all Markov chains have explored the same part of the posterior distribution \cite{gelman2013bayesian}, i.e. whether the obtained distribution is reliable or accrued due to a random effect.
    This can be monitored by convergence diagnostics such as the potential-scale-reduction statistic $\hat{R}$ \cite{gelman1992inference} and Monte Carlo standard error \cite{geyer2011introduction} (which depends on the effective sample size \cite{gelman2013bayesian}).
    Within this work, the adapted versions of these diagnostics, as proposed by Vehtari et al. \cite{vehtari2021rank}, are employed.
    In addition, trace plots \cite{gelman2011inference} can be used to monitor the behavior of chains which explore multiple marginal modes.
    For each of these diagnostics, it is essential to use multiple chains \cite{vehtari2021rank,gelman2011inference} for a reliable result.
    \\\\
    However, a multimodal posterior provides some pitfalls.
    As already mentioned at the beginning of \cref{sec:analysis_steps}, the Markov chains can get stuck in certain, isolated modes.
    Thus not all chains would have seen the same parts of the posterior distribution and the convergence diagnostics would indicate that the chains have not converged.
    Therefore, in case a multimodal posterior is studied, where all modes are of interest, the usual methods are not applicable.
    An adaptation has to be made.
    Under the assumption that all modes of the posterior were found via Monte Carlo maximum a posteriori estimation, see \cref{section:mcmle}, the following strategy is employed.
    A schematic representation of the adapted approach can be found in \cref{fig:schematic_mcmc_convergence}.
    Instead of applying the convergence diagnostics to all chains at once, the chains are clustered into groups according to their sampled parameter space and the convergence diagnostics are then applied onto each group separately\footnote{A similar approach was used in Ref.~\cite{similar_method}.}.
    Consequently, the convergence for the whole posterior is monitored.
    \\\\
    The chains can be grouped according to their similarity as follows:
    To avoid problems during the clustering process, coming from high dimensional data \cite{Jonathan99whenis}, a dimensional reduction of the chains is performed.
    Each chain, consisting of $S$ sampling points, is characterized via a vector of its quantiles, in this case the $[0.1, 0.2, 0.3, 0.4, 0.5, 0.6, 0.7, 0.8, 0.9]$ - quantiles.
    Subsequent, the corresponding distance matrix \cite{graham1971addressing} of the quantile vectors is calculated using the Euclidean metric.
    The constructed matrix serves as input for the DBSCAN algorithm \cite{ester1996density}.
    The minimal cluster size should be at least two, as this is the minimal amount of chains required to perform the $\hat{R}$ diagnostic \cite{gelman2011inference}.
    An appropriate $\epsilon$ - neighborhood for the DBSCAN algorithm can be graphically determined, for example by visualizing the Euclidean distances of the quantile vectors to each other.
    Afterwards, the correct clustering of chains can be checked visually.
    Alternatively, the two-sample Kolmogorov-Smirnov test \cite{kolmogorov1933sulla,smirnov1939estimate} or the K-Sample Anderson–Darling test \cite{doi:10.1080/01621459.1987.10478517} could be employed to compare two distributions with each other.
    \\\\
    The outlined approach still allows to adjust the number of chains $N_\text{c}$ per group and the sampling points $S$ in order to gain adequate convergence diagnostics and the desired precision for the parameter estimates.
    Within this paper, one is aiming for $\hat{R} < 1.01$ \cite{vehtari2021rank} and a relative Monte Carlo standard error in the region of a few percent.

    \subsection{Analysis of generated data}\label{sec:analysis_generated_data}
    It is crucial to prove the correct implementation and validity of the used model.
    An ideal testing scenario would be the a priori knowledge of the correct outcome of the analysis using the model under consideration.
    Therefore the PWA solution EtaMAID2018 \cite{Tiator:2018heh} is employed for the electromagnetic multipoles in \cref{eq:mulitpole_vector} up to the desired truncation order $\ell_\text{max}$.
    By these means, pseudo data for the profile functions $\check{\Omega}^{\alpha}(W, \theta)$ are generated via \cref{eq:observable} for certain energies and angular positions for the observables $\sigma_0$, $\Sigma$, $T$, $E$, $F$ and $G$.
    These data were used as input for the TPWA following the described steps in \cref{section:mcmle,sec:posteriorSampling,sec:mcmcConvergence}.
    This analysis yielded again the EtaMAID2018 multipole solutions, indicating a correct implementation.

    \section{Convergence diagnostics}\label{sec:mcmc_convergence_diagnostics}
\begin{figure*}
        \includegraphics[width=0.49\textwidth,height=\textheight,keepaspectratio]{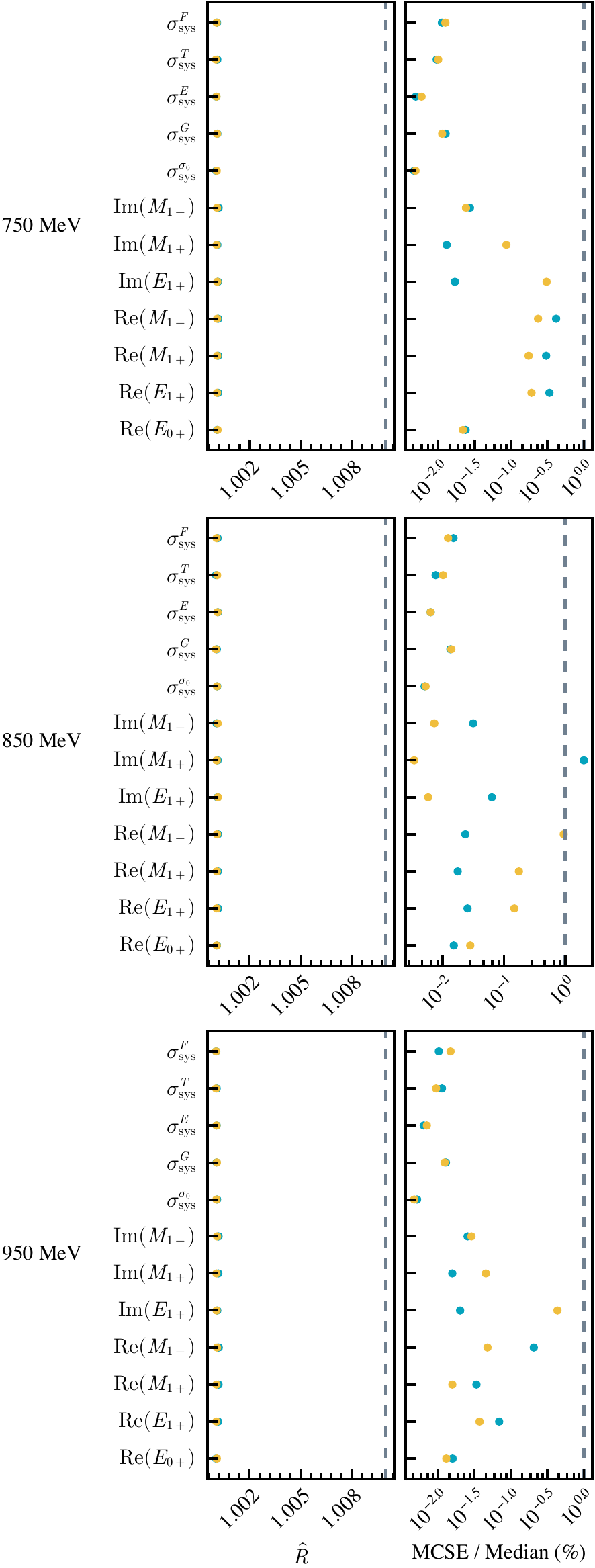}
        \includegraphics[width=0.49\textwidth,height=\textheight,keepaspectratio]{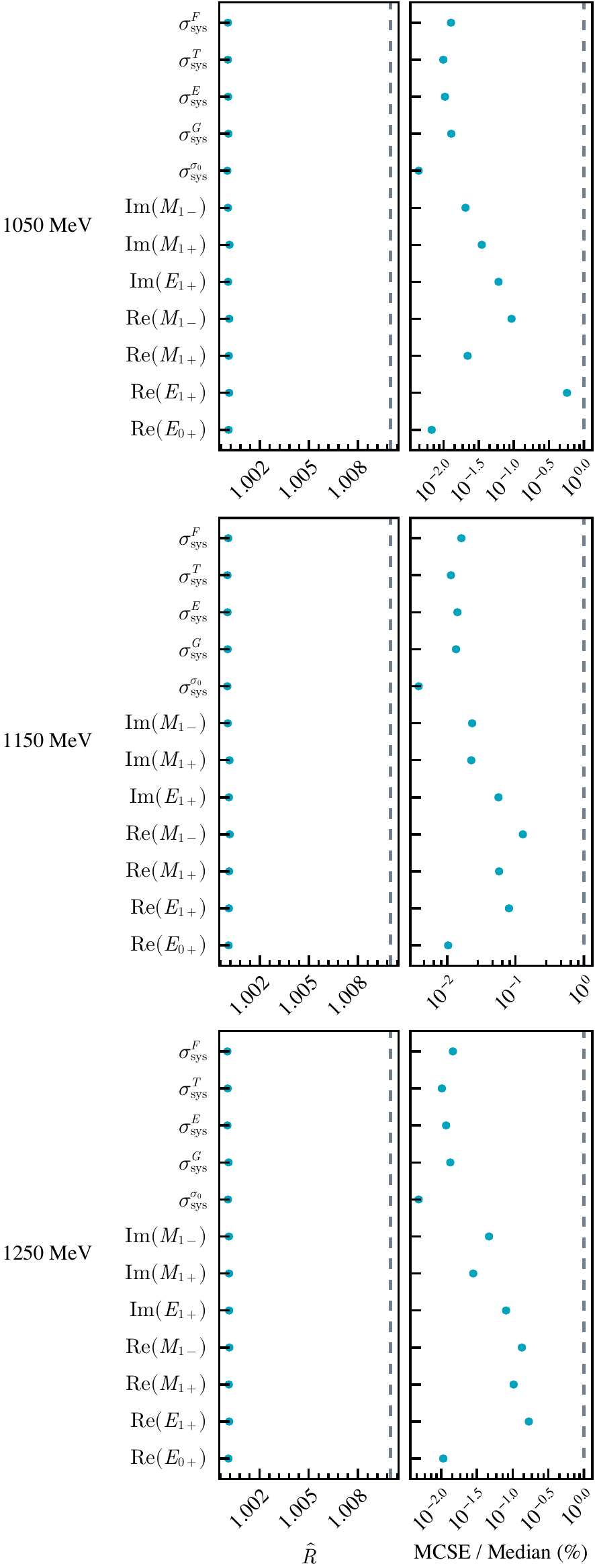}
        \caption{MCMC convergence diagnostics for the truncation order $\ell_\text{max} = 1$.
        Shown are the potential-scale-reduction statistic $\hat{R}$ (the grey, dashed line indicates the value of 1.01) and the Monte Carlo standard error (MCSE) for the median divided by the median in percent (the grey, dashed line indicates the value of $1\%$). Each solution group is drawn in a different color. \label{fig:diagnostics_lmax1}}
    \end{figure*}

\begin{figure*}
        \includegraphics[width=0.49\textwidth,height=\textheight,keepaspectratio]{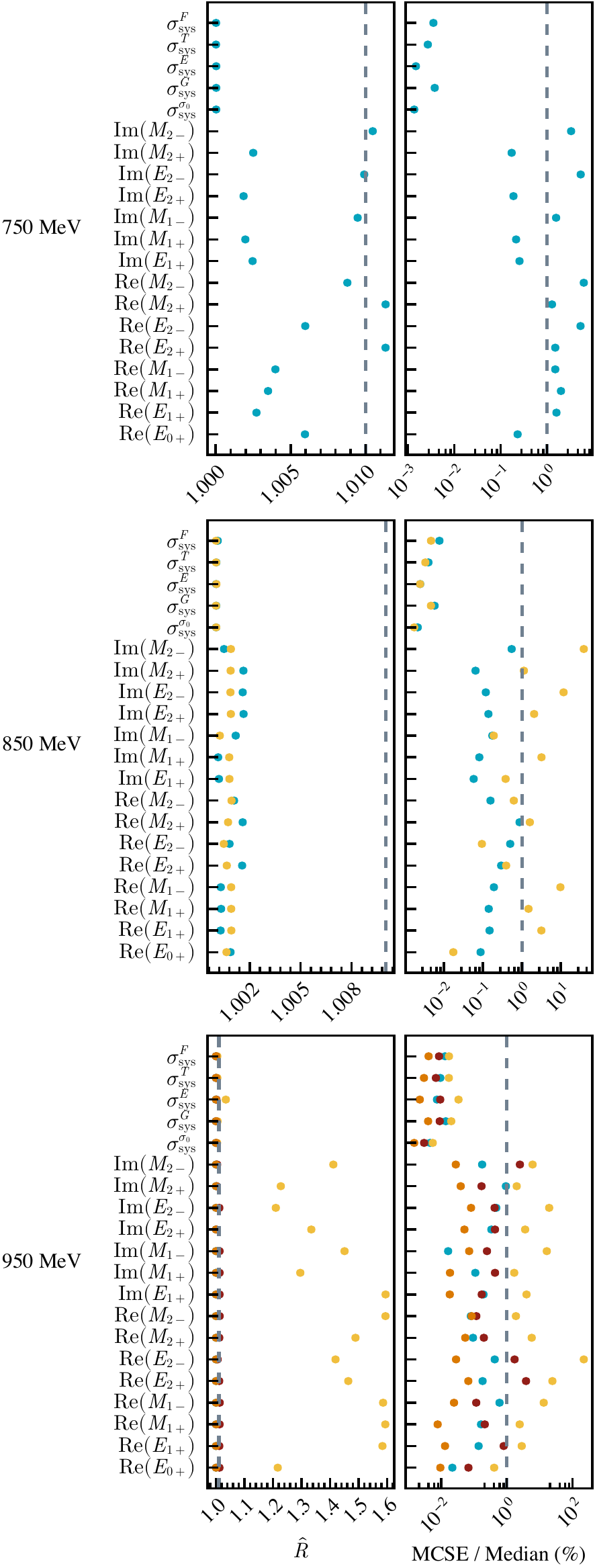}
        \includegraphics[width=0.49\textwidth,height=\textheight,keepaspectratio]{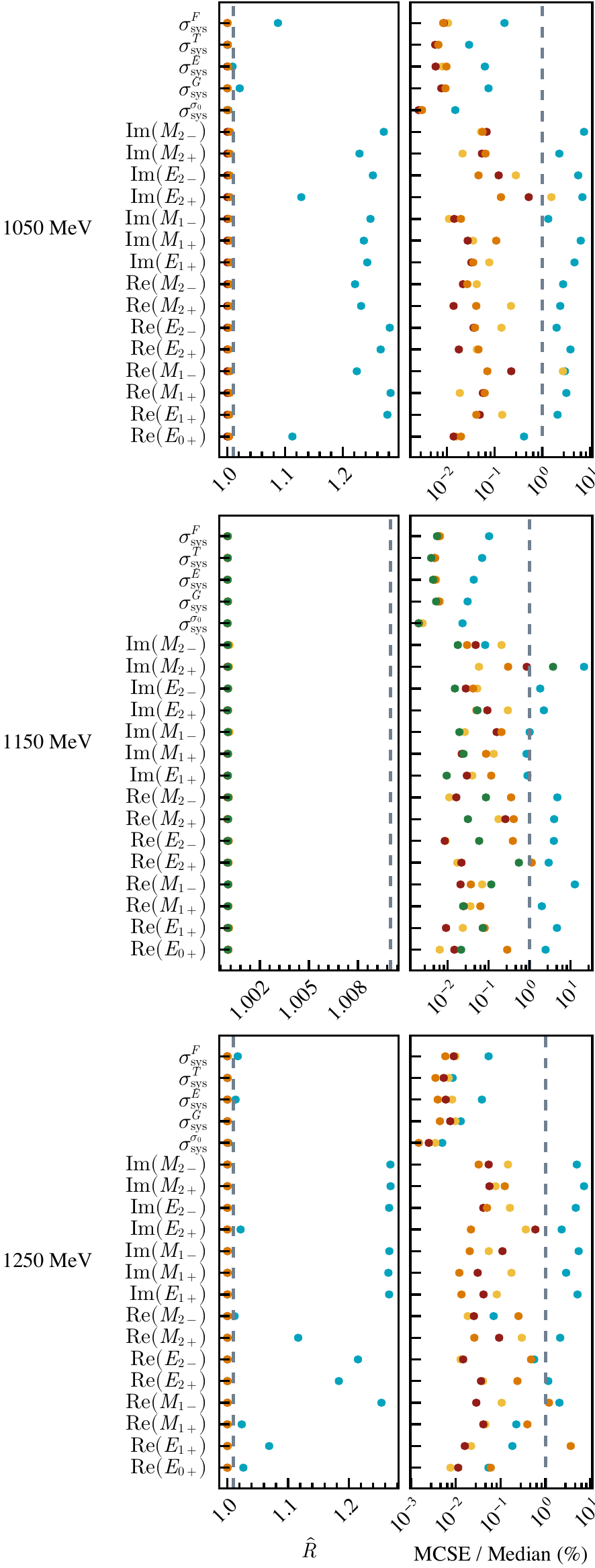}
        \caption{MCMC convergence diagnostics for the truncation order $\ell_\text{max} = 2$.
        Shown are the potential-scale-reduction statistic $\hat{R}$ (the grey, dashed line indicates the value of 1.01) and the Monte Carlo standard error (MCSE) for the median divided by the median in percent (the grey, dashed line indicates the value of $1\%$). Each solution group is drawn in a different color. \label{fig:diagnostics_lmax2}}
    \end{figure*}

    MCMC convergence diagnostics for the truncation orders $\ell_\text{max} = 1$ and $\ell_\text{max} = 2$ for all analyzed energies are shown in \cref{fig:diagnostics_lmax1,fig:diagnostics_lmax2}.
    The anticipated values for the potential-scale-reduction statistic $\hat{R} \le 1.01$ and the relative Monte Carlo standard error (MCSE) of the median in the range of a few percent were achieved for both truncation orders.
    The diagnostics for $750$~MeV are satisfactory, despite their slightly elevated values, which are the result of the highly multi-modal marginal parameter distribution.
    However, certain convergence diagnostics for $\ell_\text{max} = 2$ suggest that some groups of chains have not yet converged, indicating a specific phenomenon that will be discussed below.
    Hence, the four energies $E^\text{lab}_{\gamma}=[950, 1050, 1150, 1250]$~MeV look suspicious.
    In each case one group of chains show $\hat{R}$-values way above 1.01 and relative Monte Carlo standard errors of over $100\%$.
    This results from two modes separated in phase space by a small region of low probability, so that the Metropolis acceptance probability \cite{neal2011mcmc} for a transition between the two high probability regions is quite small but nonzero.
    Hence, just a small number of chains is able to explore both marginal modes at once, which is the reason for the suspicious convergence diagnostics.
    For the case of $1050$~MeV, the blue distribution corresponds to a cluster with just one group member.
    Hence, it is not possible to calculate an $\hat{R}$-value for this cluster.
    It is important to note that this behavior can not be prevented as it is inherently a random effect.
    As an example how such a phenomenon manifests within a parameter distribution, see the blue distribution of $\Im(M_{2+})$ at $1250 \; \mathrm{MeV}$ in \cref{fig::multipole_plots_lmax_2_m2p_2}.
    Despite their convergence diagnostics, these types of distributions are shown within the multipole parameter and posterior predictive plots for their illustrative purposes.

    \section{Probability distributions for the quotient and product of two Gaussian random-variables}\label{sec:ProbDistsDetails}
    Assuming the original observables to follow a Gaussian probability distribution up to a very good approximation, the result of forming the quotient and/or product is generally non-Gaussian.
    This appendix collects some basic facts about the quotient- and the product-distribution and considers some limiting cases.

    \subsection{The quotient distribution: $Z := X / Y$}\label{sec:QuotientDistribution}
    Given are two independent, uncorrelated, Gaussian distributed random variables $X$ and $Y$:
    \begin{equation}
        X \sim \mathcal{N} \left( \mu_X, \sigma_X \right) \mathrm{,} \hspace*{5pt} Y \sim \mathcal{N} \left( \mu_Y, \sigma_Y \right) \mathrm{,} \label{eq:TwoUncorrelatedNormalRandomVars}
    \end{equation}
    together with the integral defining the probability distribution function of the quotient variable $Z := X / Y$ \cite{definition_ratio_normal}:
    \begin{align}
        \mathrm{P}_{X / Y} (u) &= \int_{-\infty}^ {\infty} \dd{x} \int_{-\infty}^ {\infty} \dd{y} \delta \left(\frac{x}{y} - u \right) \notag \\
        & \times \frac{\exp\qty[-\frac{1}{2}\qty(\frac{( x - \mu_X )^2}{\sigma_X^{2}} + \frac{\left( y - \mu_Y \right)^{2}}{\sigma_Y^{2}})]}{2 \pi \sigma_X \sigma_Y} \notag \\
        &= \int_{-\infty}^{\infty} \dd{y} \left| y \right| \frac{\exp\qty[-\frac{1}{2}\qty(\frac{( uy - \mu_X )^2}{\sigma_X^{2}} + \frac{\left( y - \mu_Y \right)^{2}}{\sigma_Y^{2}})]}{2 \pi \sigma_X \sigma_Y}. \label{eq:QuotientPDFGeneralExpression}
    \end{align}
    Mathematica yields the following result (for positive values of~$\sigma_X$ and~$\sigma_Y$):
    \begin{align}
        \mathrm{P}_{X / Y}(u) &= f_1(u) \cdot \left[\sqrt{2} \; f_2(u) \; c \right. \notag\\
        & \left. + \sqrt{\pi} \; f_3(u) \; \text{erf}\qty(f_4(u)) \exp\qty(f_4(u)^2) \right], \label{eq:QuotientIntClosedExpressionMathematica}
    \end{align}
    with the declarations:
    \begin{align}
        f_1(u) &:= \frac{\exp(-\frac{1}{2}\qty(\frac{\mu_X^2}{\sigma_X^2} + \frac{\mu_Y^2}{\sigma_Y^2}))}{\sqrt{2} \pi f_2(u)^3},\\
        f_2(u) &:= \sqrt{\sigma_X^2+\sigma_Y^2 u^2}, \\
        f_3(u) &:= \mu_Y \sigma_X^2+\mu_X \sigma_Y^2 u, \\
        f_4(u) &:= \frac{f_3(u)}{\sqrt{2} \; f_2(u) \; c}, \\
        c &:= \sigma_X \sigma_Y
    \end{align}
    and the error function 'erf' \cite{definition_error_function}.
    In the following, two limiting cases for \cref{eq:QuotientIntClosedExpressionMathematica} are analyzed:
    first, the vanishing of the expectation values (i.e. $\mu_X = \mu_Y = 0$):
    \begin{equation}
        \mathrm{P}_{X / Y} (u) = \frac{\sigma_X \sigma_Y}{\pi  \qty(\sigma_X^2+\sigma_Y^2 u^2)}. \label{eq:SpecialCaseI}
    \end{equation}
    This is a result which is known from earlier publications on the quotient distribution, for instance \cite{Curtiss1941}.

    Second, considering also unit standard-deviations (i.e. $\sigma_X=\sigma_Y=1$) the result \cref{eq:SpecialCaseI} further simplifies to:
    \begin{equation}
        \mathrm{P}_{X / Y} (u)  =  \frac{1}{\pi  \left( 1 + u^2\right)}. \label{eq:SpecialCaseII}
    \end{equation}
    This is the well-known Cauchy distribution.

    \subsection{The product distribution:~$Z := X Y$}\label{sec:ProductDistribution}
    Similar to \cref{eq:QuotientPDFGeneralExpression} the probability-distribution function for the product of two independent, uncorrelated Gaussian distributed random variables can be written \cite{definition_prod_normal}:
    \begin{align}
        \mathrm{P}_{X  Y} (u) &= \int_{- \infty}^ {\infty} \dd{x} \int_{-\infty}^{\infty} \dd{y} \delta \qty( x y - u) \notag \\
        & \times \frac{\exp\qty[-\frac{1}{2}\qty(\frac{( x - \mu_X )^2}{\sigma_X^{2}} + \frac{\left( y - \mu_Y \right)^{2}}{\sigma_Y^{2}})]}{2 \pi \sigma_X \sigma_Y}. \label{eq:ProductPDFGeneralExpression}
    \end{align}
    By introducing an integral-representation for the $\delta$-function
    \begin{equation}
        \delta \qty(x y - u) = \int_{-\infty}^{+\infty} \frac{\dd{k}}{2 \pi} e^{i k \qty( x y - u)} = \int_{- \infty}^{+ \infty} \frac{\dd{k}}{2 \pi} e^{i k x y } e^{- i k u},
    \end{equation}
    one can bring \cref{eq:ProductPDFGeneralExpression} into the following form:
    \begin{equation}
        \mathrm{P}_{X  Y} (u) = \int_{-\infty}^{+\infty} \frac{\dd{k}}{2 \pi} e^{- i k u} F_{k} \qty[\mu_X, \sigma_X ; \mu_Y, \sigma_Y], \label{eq:ProductPDFGeneralIntStepI}
    \end{equation}
    where
    \begin{align}
        & F_{k} \qty[ \mu_X, \sigma_X ; \mu_Y, \sigma_Y] = \int_{-\infty}^ {\infty} \dd{x} \int_{-\infty}^ {\infty} \dd{y} e^{i k x y } \notag \\
        & \times \frac{\exp\qty[-\frac{1}{2}\qty(\frac{( x - \mu_X )^2}{\sigma_X^{2}} + \frac{\left( y - \mu_Y \right)^{2}}{\sigma_Y^{2}})]}{2 \pi \sigma_X \sigma_Y}.
    \end{align}
    This characteristic function can be solved analytically:
    \begin{align}
        & F_{k} \qty[ \mu_X, \sigma_X ; \mu_Y, \sigma_Y] \notag \\
        & = \int_{-\infty}^{\infty} \dd{y} \exp \qty[-\frac{1}{2} k y \qty(k \sigma_X^2 y-2 i \mu_X)] \frac{\exp \left[-\frac{\left( y - \mu_Y \right)^{2}}{2 \sigma_Y^{2}}\right]}{\sqrt{2 \pi} \sigma_Y}  \nonumber \\
        & = \frac{  \exp \left[-\frac{k \left(k \mu_Y^2 \sigma_X^2+k \mu_X^2 \sigma_Y^2-2 i \mu_X \mu_Y\right)}{2 + 2 k^2 \sigma_X^2 \sigma_Y^2 }\right]}{ \sqrt{ 1 + k^2 \sigma_X^2 \sigma_Y^2}} . \label{eq:FourierCoeffSecondIntDone}
    \end{align}
    The final result has the shape of a Fourier integral:
    \begin{align}
        \mathrm{P}_{X  Y} (u) &=  \int_{- \infty}^{+ \infty} \frac{d k}{2 \pi} \exp \left[- i k u \right] \nonumber \\
        & \times \frac{  \exp \left[-\frac{k \left(k \mu_Y^2 \sigma_X^2+k \mu_X^2 \sigma_Y^2-2 i \mu_X \mu_Y\right)}{2 + 2 k^2 \sigma_X^2 \sigma_Y^2 }\right]}{ \sqrt{ 1 + k^2 \sigma_X^2 \sigma_Y^2}}  . \label{eq:ProductDistFourierFinalResult}
    \end{align}
    In analogy to the quotient distribution, the limiting case $\mu_X = \mu_Y = 0$ shall be analyzed.
    The Fourier coefficients become:
    \begin{equation}
        F_{k} \left[ 0, \sigma_X ; 0, \sigma_Y \right] = \frac{1}{\sqrt{ 1 + k^2 \sigma_X^2 \sigma_Y^2}}. \label{eq:FourierCoeffBothMeansZero}
    \end{equation}
    The result for the product distribution can in this case be written with a modified Bessel function of the second kind~$K_{n} (z)$:
    \begin{equation}
        \mathrm{P}_{X  Y} (u) = \frac{K_0\left(\frac{| u| }{\sigma_X \sigma_Y}\right)}{\pi  \sigma_X \sigma_Y}. \label{eq:PDFResultBothMeansVanish}
    \end{equation}
    This is the analogue of \cref{eq:SpecialCaseI} from the case of the quotient distribution.
    For unit standard deviations, \cref{eq:PDFResultBothMeansVanish} becomes simply $K_{0} \qty(\left| u \right|) / \pi$, which is the analogue of \cref{eq:SpecialCaseII}.

    For the product distribution, especially one limiting case is of interest for this paper, namely where the standard deviation of one random variable almost vanishes (i.e. $\sigma_Y \rightarrow 0$).
    The characteristic function becomes:
    \begin{align}
        \lim_{\sigma_Y \to 0} F_{k} = \exp \qty(-\frac{k \qty(k \mu_Y^2 \sigma_X^2 - 2 i \mu_X \mu_Y)}{2}). \label{eq:FourierCoeffSigma2Vanishes}
    \end{align}
    Substituting \cref{eq:FourierCoeffSigma2Vanishes} into \cref{eq:ProductDistFourierFinalResult} and solving the integral gives the result:
    \begin{equation}
        \mathrm{P}_{X  Y} (u) = \frac{\exp\qty(-\frac{\qty(u-\mu_X \mu_Y)^2}{2 \mu_Y^2 \sigma_X^2})}{\sqrt{2 \pi } \qty| \mu_Y|  \qty| \sigma_X| }, \label{eq:PDFResultSigma2Vanishes}
    \end{equation}
    which is indeed a Gaussian probability distribution function.
    This result is used in \cref{sec:assumptions}.

    \clearpage
    \bibliography{paper}

\begin{thebibliography}{84}
\bibitem [{sup()}]{supplementMaterial}%
  \BibitemOpen
  \href@noop {} {\bibinfo {title} {See {S}upplemental {M}aterial at [url will
  be inserted by publisher] for all parameter figures and the implementation of
  the used model within {S}tan.}}\BibitemShut {Stop}%
\end{thebibliography}

    \end{document}